\documentclass[12pt]{article}

\usepackage[margin=1in]{geometry}

\usepackage{setspace}

\usepackage{graphicx}

\usepackage{enumitem}

\usepackage{microtype}
\usepackage{subfigure}
\usepackage{booktabs}
\usepackage{comment}
\usepackage{wrapfig}
\usepackage{arydshln}
\usepackage{enumitem}
\usepackage{multirow}
\usepackage{url}
\usepackage{natbib}
\usepackage{authblk}

\RequirePackage[colorlinks,citecolor=blue,linkcolor=blue,urlcolor=blue,pagebackref]{hyperref}

\usepackage{amsmath}
\usepackage{amssymb}
\usepackage{mathtools}
\usepackage{amsfonts}
\usepackage{bm}
\usepackage{bbm}

\usepackage{algorithm}
\usepackage{algorithmic}

\def\1{\bm{1}}

\DeclareMathAlphabet{\mathsfit}{\encodingdefault}{\sfdefault}{m}{sl}
\SetMathAlphabet{\mathsfit}{bold}{\encodingdefault}{\sfdefault}{bx}{n}

\usepackage{amsthm}

\theoremstyle{plain}

\usepackage[textsize=tiny]{todonotes}
\usepackage{multirow}
\usepackage{wrapfig}
\usepackage{subfigure}
\usepackage{tabularx}

\usepackage{xcolor}
\newcount\Comments  % 0 suppresses notes to selves in text
\newcommand{\kibitz}[2]{\ifnum\Comments=1\textcolor{#1}{#2}\fi}

\usepackage{listings}

\usepackage{tabularx}

\usepackage{tikz}
\usetikzlibrary{arrows.meta,positioning,fit,calc,shapes.geometric}

\newcolumntype{Y}{>{\raggedright\arraybackslash}X}
\newcolumntype{C}{>{\centering\arraybackslash}X}

\newcommand{\ModelRun}{\texttt{ModelRun}}
\newcommand{\Relation}{\texttt{Relation}}
\newcommand{\ObservedValue}{\texttt{ObservedValue}}
\newcommand{\ModelOutputPoint}{\texttt{ModelOutputPoint}}

\newcommand{\ReportSentence}{\texttt{ReportSentence}}
\newcommand{\OutputTest}{\texttt{OutputTest}}

\usepackage {framed}

\hypersetup{
  pdftitle={AI Economist Agent: An Agentic Framework for Evidence-Based Economic and Financial Analysis with RAG, Knowledge Graphs, and Large Language Models},
  pdfauthor={Masahiro Kato},
  pdfsubject={Economic scenario analysis with GraphRAG, explicit model execution, stress testing, and source traceability},
  pdfkeywords={large language models, GraphRAG, financial modeling, scenario testing, source traceability, model validation}
}
\newcommand{\GPTFourOSnapshot}{\texttt{gpt-4o-\allowbreak{}2024-\allowbreak{}08-\allowbreak{}06}}

\usepackage[capitalize,noabbrev]{cleveref}

\allowdisplaybreaks

\title{AI Economist Agent: An Agentic Framework for Evidence-Based Economic and Financial Analysis with RAG, Knowledge Graphs, and Large Language Models}

\author{Masahiro Kato\thanks{Email: \texttt{mkato-csecon@g.ecc.u-tokyo.ac.jp}}$\,$}

\affil{The University of Tokyo}

\date{\today}
\begin{document}

\maketitle

\begin{abstract}
We propose an AI economist agent for economic and financial scenario analysis. Scenario design often requires analysts to assess emerging risks with limited historical precedent, combine information from many sources, and translate qualitative mechanisms into internally consistent quantitative paths. Large language models (LLMs) can search and synthesize this information, but fluent narratives alone do not establish the model-based calculations needed for economic conclusions. Our framework uses LLM agents to plan the analysis, retrieve relevant evidence, and organize economic mechanisms, while registered quantitative models generate numerical outcomes and predefined tests determine whether intermediate results can be used in the final report. We apply the framework to European macro-financial stress scenarios and bank capital analysis. The empirical analysis evaluates retrieval of economic mechanisms, scenario construction, model execution, and report generation under a historical information cutoff. The results show how the AI economist agent can combine flexible evidence retrieval and scenario construction while keeping the resulting analysis linked to identifiable sources and explicit model calculations.
\end{abstract}


\section{Introduction}
\label{sec:introduction}

Economic and financial decisions often depend on risks for which historical precedent is limited. A new geopolitical or financial disturbance can propagate through markets, firms, households, and financial institutions in combinations that are not represented by a single past episode. Scenario analysis must therefore remain flexible enough to examine new combinations of shocks while retaining the economic mechanisms and quantitative assumptions that determine the results.

Large language models (LLMs) are useful in this setting because they can search and synthesize large collections of economic documents. Their flexibility, however, creates a basic problem for economic analysis. A plausible narrative can combine statements that come from different sources without preserving the conditions under which those statements were made, and it can state numerical consequences without performing the model calculation needed to support them. Economic analysis requires both forms of discipline: qualitative claims must remain connected to identifiable evidence, and quantitative claims must be obtained from an explicit model with defined inputs and outputs.

We propose an AI economist agent that combines information retrieval, a graph of economic relations, and quantitative model execution in one framework. The LLM agents plan the analysis, retrieve relevant information, organize transmission mechanisms, and formulate structured requests to the available models. Numerical conclusions are produced by registered quantitative models rather than by the LLM alone, and predefined checks determine whether each result can be used in the report. This division of tasks uses the flexibility of LLMs while retaining the structure required for economic analysis.

The framework is designed for applications in which the relevant evidence changes with the question and the date of analysis. The same structure can support macroeconomic, investment, corporate, and financial stability analysis when the task requires both documentary evidence and explicit calculation. We study a European stress scenario because it requires the system to identify risks from dated information, trace their transmission through the economy, generate joint macro-financial paths, and translate those paths into bank capital outcomes.

\subsection{Contribution}

Our main contribution is to formalize and implement an algorithm for economic scenario generation with LLM agents. The algorithm starts from an analysis request and an admissible information set, retrieves evidence relevant to the question, organizes that evidence into economic transmission channels, formulates a structured model request, executes compatible quantitative models, and generates a report from the accepted results. The implementation records the objects created at each stage so that the analysis can be inspected after execution.

Three design choices are central to the algorithm. The knowledge base retains the source passage and scope of extracted economic relations and supports both document retrieval and graph search. Quantitative claims require an executed model from a registered set of specifications, so a narrative estimate cannot substitute for a calculation. Acceptance tests are fixed before the corresponding outputs are observed and determine whether evidence, model results, and reported numbers can pass to the next stage. This use of tests adapts the logic of test-driven development (TDD) to the interfaces of an economic analysis \citep{Beck2002testdriven}.

We implement the framework in European stress scenario design. The empirical analysis evaluates document and graph retrieval, construction of the relation graph, model-based scenario generation, bank capital calculations, and execution of the agent program. These experiments examine whether the framework can organize economic evidence, invoke quantitative models, and preserve the records needed to inspect the resulting analysis.

The term \emph{AI economist agent} refers to this allocation of analytical tasks. It is distinct from the AI Economist of \citet{Zheng2022theai}, which studies tax policy design through multi-agent reinforcement learning. Our framework is designed to support economic analysis by linking retrieved information to explicit calculations and preserving the records used to produce the report.

\subsection{Related Work}

This study builds on work on AI agents for economic research, retrieval-augmented generation, knowledge graphs, and quantitative scenario analysis. LLM agents can coordinate research tasks and tool use \citep{Korinek2023generativeai,Korinek2025aiagents}, while retrieval-augmented generation and graph-based retrieval provide different ways to organize information distributed across documents \citep{Lewis2020retrievalaugmented,Edge2025fromlocal}. We use these methods to support economic analysis rather than to replace the quantitative models that determine numerical outcomes.

The construction of the relation graph also draws on work that decomposes source text before extracting structured facts. In particular, ATOM motivates the use of short statements when building the relation database from economic documents \citep{Lairgi2026atomadaptive}. Conditional forecasting and structural restrictions provide the quantitative basis for converting qualitative economic restrictions into simulated paths \citep{Waggoner1999conditionalforecasts,Banbura2015conditioalforecasts,AntolinDiaz2018narrativesign}. The literature on stress testing motivates the separation between scenario design and the translation of scenarios into bank outcomes \citep{BCBS2018stresstesting,Covas2014stresstestingus,Parlatore2025designingstress,Aikman2024stresstesing}. The Appendix provides a fuller discussion of these literatures.

The remainder of this study is organized as follows. Section~2 defines the analysis request and the information available to the agent. Section~3 presents the AI economist agent, its components, algorithm, construction of the knowledge base, and acceptance tests. Section~4 applies the framework to stress scenario design. Section~5 reports the empirical analysis, and Section~6 concludes.

\section{Setup}
\label{sec:setup}

This section defines the objects that are fixed before the AI economist agent begins an analysis. The setup consists of the analysis request and the information available for that request.

\subsection{Analysis Request}

We represent an analysis request by
\begin{align}
 q=\left(a_q,\mathcal{J}_q,\mathcal{K}_q,H_q,\mathcal{O}_q,\mathcal{C}_q\right),
 \label{eq:analysisrequest}
\end{align}
where $a_q$ identifies the application, $\mathcal{J}_q$ and $\mathcal{K}_q$ identify the jurisdictions and sectors, $H_q$ is the horizon, and $\mathcal{O}_q$ contains the requested outcomes. The final element $\mathcal{C}_q$ contains restrictions on admissible sources, variables, models, and data. A historical analysis, for example, can require that every source was public before a specified date. An investment application can require exposures for individual securities and valuation outputs, while a bank application can require exposure data for individual institutions and Common Equity Tier 1 (CET1) outcomes.

\subsection{Information Available to the Agent}

The information available for request $q$ is
\begin{align}
 \mathcal{I}_q=\left(\mathcal{S}_q,\mathcal{N}_q,\mathfrak{M}_q\right),
 \label{eq:informationset}
\end{align}
where $\mathcal{S}_q$ contains retrievable sources, $\mathcal{N}_q$ contains observed values, and $\mathfrak{M}_q$ contains the quantitative model specifications available for the request. The sources may include narrative documents, structured tables, data releases, and model documentation. The model specifications are stored separately because a statement about an economic relation and a model that computes an outcome play different roles in the analysis.

Each source location identifies where a relation or value was obtained. A prose location records the file and quoted passage, a spreadsheet location records the workbook and cell coordinates, and a database series is identified by its provider and series metadata. The information date and other source restrictions in $\mathcal{C}_q$ determine which of these locations may be used for a particular analysis.

Observed values are represented by
\begin{align}
 w=\left(v_w,x_w,u_w,f_w,j_w,k_w,e_w,\mu_w\right),
 \label{eq:observedvalue}
\end{align}
where $v_w$ identifies the variable, $x_w$ is its value, $u_w$ is the unit, $f_w$ is the frequency, and $j_w$ and $k_w$ identify the jurisdiction and sector. The source location is $e_w$, while $\mu_w$ contains information required by the application, such as the reference period, release date, data vintage, or scenario condition. An observed value can be supplied to a model only when its variable definition, unit, frequency, scope, and required metadata agree with that model specification.

\section{The AI Economist Agent}
\label{sec:agent}

The AI economist agent transforms the request $q$ and information set $\mathcal{I}_q$ into an economic analysis. LLM agents handle planning and interpretation, while retrieval, model execution, and testing follow explicit program rules. After an overview, this section defines the main components and algorithm, explains how the knowledge base is constructed, and describes the acceptance tests used by the algorithm.

\subsection{Overview}

The analysis begins by identifying the variables and outcomes required by the request. The agent retrieves eligible source material, searches for economic relations that connect the initiating condition to the requested outcomes, and organizes those relations into candidate channels. The selected channels identify the variables and restrictions used to form a structured model request. The execution layer runs the compatible registered models and stores their outputs, after which the accepted records are used to generate the report.

Figure~\ref{fig:agent-structure} summarizes this sequence. Source preparation occurs before an individual analysis, while retrieval, model execution, and reporting are performed for each request. The same sequence can be used with different economic applications because the request determines the variables, sources, models, and outcomes that are relevant to the analysis.

\begin{figure}[!tbp]
\centering
\begin{tikzpicture}[
 node distance=1.55cm and 0.85cm,
 box/.style={draw,rounded corners,align=center,minimum height=0.90cm,text width=2.75cm},
 arr/.style={-{Latex[length=2mm]},thick}
]
\node[box] (sources) {Text, tables, numerical series, and model documentation};
\node[box,right=of sources] (parser) {Source parsing and validation};
\node[box,right=of parser] (stores) {Retrieval indexes, knowledge graph, and model specifications};

\node[box,below=of sources] (retrieve) {Retrieval of eligible sources};
\node[box,right=of retrieve] (graph) {Search for economic channels};
\node[box,right=of graph] (request) {Variables, restrictions, and required outputs};

\node[box,below=of request] (model) {Execution of compatible quantitative models};
\node[box,left=of model] (application) {Application calculations};
\node[box,left=of application] (selection) {Selection of candidates};

\node[box,below=of selection] (tests) {Acceptance tests};
\node[box,right=of tests] (report) {Report generation};

\draw[arr] (sources) -- (parser);
\draw[arr] (parser) -- (stores);
\draw[arr] (stores.south) -- ++(0,-0.35) -| (retrieve.north);
\draw[arr] (retrieve) -- (graph);
\draw[arr] (graph) -- (request);
\draw[arr] (request) -- (model);
\draw[arr] (model) -- (application);
\draw[arr] (application) -- (selection);
\draw[arr] (selection) -- (tests);
\draw[arr] (tests) -- (report);
\end{tikzpicture}
\caption{AI economist agent. Source material is prepared before an analysis begins. For each request, the agent retrieves evidence and economic relations, submits a structured request to compatible quantitative models, applies the required tests, and generates a report from the accepted records.}
\label{fig:agent-structure}
\end{figure}

\subsection{Components}

The framework uses economic relations and channels to represent the substantive content of the analysis, a knowledge graph to organize the relevant records, and quantitative models to compute numerical outcomes. These objects have separate roles even when they refer to the same economic variables.

\paragraph{Economic relations and channels.}
For each eligible source, the knowledge base can store an economic relation $c$ as
\begin{align}
 c=\left(v_c^{\mathrm{src}},v_c^{\mathrm{dst}},\kappa_c,\sigma_c,h_c,j_c,k_c,g_c,m_c,e_c,\mu_c\right),
 \label{eq:sourcerelation}
\end{align}
where $v_c^{\mathrm{src}}$ and $v_c^{\mathrm{dst}}$ are variables from a controlled vocabulary. The vocabulary maps synonymous expressions, such as ``policy rate'' and ``short-term policy interest rate,'' to one variable identifier. The evidence class $\kappa_c$ belongs to
\begin{align}
\begin{aligned}
 \mathcal{R}_c=\{&\text{causal},\text{structural},\text{predictive},\text{accounting},\\
 &\text{institutional},\text{descriptive},\text{policy judgment}\}.
\end{aligned}
 \label{eq:relationclasses}
\end{align}
The sign $\sigma_c$ records whether the relation is positive, negative, non-monotone, ambiguous, or absent. Under a positive sign, an increase in the source variable raises the target variable; under a negative sign, it lowers the target variable. Here, $h_c$ is the horizon, $j_c$ and $k_c$ identify the jurisdiction and sector, $g_c$ records the economic regime, $m_c$ records the empirical basis, $e_c$ identifies the exact source location, and $\mu_c$ contains any additional information required by the application. We record the evidence class separately from the relation verb assigned during extraction because a directional verb does not by itself make a descriptive association causal evidence. A descriptive relation can provide context for the report, but it may not be relabeled as a causal estimate.

An economic channel is an ordered sequence of these relations that connects an initiating condition to an outcome. We represent one channel as
\begin{align}
 C=\left(v_0,r_1,v_1,r_2,\ldots,r_L,v_L\right),
 \label{eq:channel}
\end{align}
where relation $r_{\ell}$ connects variable $v_{\ell-1}$ to variable $v_{\ell}$ for $\ell=1,\ldots,L$. The initial variable $v_0$ identifies the condition from which the analysis begins, and $v_L$ is the target outcome. Because one initiating event can affect several outcomes through different mechanisms, scenario $s$ can contain a set of accepted channels,
\begin{align}
 \mathcal{C}_s=\{C_{s,1},C_{s,2},\ldots,C_{s,K_s}\}.
 \label{eq:channelset}
\end{align}
Keeping the channels separate makes it possible to identify which evidence and model inputs support each part of the analysis. A policy rate change, for example, can affect investment and output through borrowing costs, bond portfolios through the yield curve, and bank income through lending conditions. The relations in one channel need not share an evidentiary status: a channel can contain a causal estimate, a structural relation, an accounting identity, or an institutional rule, provided that the role of each relation is stated separately.

Figure~\ref{fig:economic-channels} illustrates this representation.

\begin{figure}[!htbp]
\centering
\begin{tikzpicture}[
 node distance=1.15cm and 1.15cm,
 box/.style={draw,rounded corners,align=center,minimum height=0.85cm,text width=2.45cm},
 arr/.style={-{Latex[length=2mm]},thick}
]
\node[box] (change) {Policy rate change};
\node[box,right=of change] (borrowing) {Borrowing costs};
\node[box,right=of borrowing] (investment) {Investment and output};
\node[box,below=of borrowing] (yields) {Yield curve};
\node[box,right=of yields] (bonds) {Bond portfolio value};
\node[box,above=of borrowing] (lending) {Lending conditions};
\node[box,right=of lending] (income) {Bank income and credit risk};
\draw[arr] (change) -- (borrowing);
\draw[arr] (borrowing) -- (investment);
\draw[arr] (change) |- (yields);
\draw[arr] (yields) -- (bonds);
\draw[arr] (change) |- (lending);
\draw[arr] (lending) -- (income);
\end{tikzpicture}
\caption{Illustrative economic channels from a policy rate change to macroeconomic, investment, and banking outcomes. Each arrow represents a relation supported by an eligible source, an accounting identity, or a model equation.}
\label{fig:economic-channels}
\end{figure}

\paragraph{Knowledge graph and retrieval.}
The knowledge graph is a directed graph $G=(\mathcal{N},\mathcal{A})$, where $\mathcal{N}$ is the set of nodes and $\mathcal{A}$ is the set of directed links. It stores source locations, economic relations, observed values, model specifications, calculation records, and reporting records as separate objects. Co-occurrence of two variables in one passage does not create a relation. A \Relation{} links the exact source location to the variables and relation fields in Equation~\ref{eq:sourcerelation}, while an \ObservedValue{} links a numerical observation to its definition and source. A value produced by a model is stored as a \ModelOutputPoint{} linked to the \ModelRun{} that produced it: a relation reported in a source is not a numerical observation, an observed value does not establish why two variables are related, and a calculated value depends on the full calculation record.

\begin{table}[!htbp]
\centering
\caption{Principal objects stored in the knowledge graph.}
\label{tab:graph-objects}
\begin{tabularx}{\linewidth}{p{0.23\linewidth}p{0.38\linewidth}Y}
\toprule
Layer & Principal objects & Role \\
\midrule
Source & \texttt{Source}, \texttt{SourceLocation} & Retain evidence locations \\
Economic relation & \Relation{} & Record interpreted links \\
Observed data & \ObservedValue{} & Identify input values \\
Economic concepts & \texttt{Variable}, \texttt{Jurisdiction} & Define variable and scope \\
Quantitative model & \texttt{ModelSpecification} & Define executable calculation \\
Calculation & \ModelRun{}, \ModelOutputPoint{} & Retain numerical results \\
Reporting & \ReportSentence{}, \OutputTest{} & Link claims to their tests \\
\bottomrule
\end{tabularx}
\end{table}

Document RAG retrieves passages from the eligible source collection, while GraphRAG searches the directed relation graph for paths relevant to the request. Both methods retain the underlying source locations. Graph retrieval adds the ordering of relations needed to form an economic channel and shows where a proposed transmission mechanism has no connecting relation. Admissible paths are ranked by a fixed score that uses the extraction confidence, the number of attached source documents, and the path length; the score is stated in the Appendix. Two paths are grouped only when their ordered variables and relation signs agree, so a different sign pattern remains a distinct candidate. Records of model execution are append-only: each \ModelRun{} has a unique identifier and stores its specification and request together with the hashes, timing, environment, and diagnostics needed to reproduce the calculation, so a report sentence can be traced through the calculation and its restrictions to the underlying economic relation and source location.

\paragraph{Quantitative models.}
The agent can call only quantitative models whose specifications belong to $\mathfrak{M}_q$ in Equation~\ref{eq:informationset}. Each specification defines its inputs and outputs together with their units, frequency, horizon, scope, estimation or calibration snapshot, and executable implementation. To prevent selection of a model after its results are observed, each empirical application records the model specifications used for the comparison before calculating the corresponding outputs.

\begin{table}[!htbp]
\centering
\caption{Examples of quantitative models available to the agent.}
\label{tab:model-blocks}
\begin{tabularx}{\linewidth}{p{0.26\linewidth}Yp{0.28\linewidth}}
\toprule
Application & Example model & Output \\
\midrule
Macroeconomic scenario & Vector autoregression & Joint variable paths \\
Investment valuation & Asset pricing model & Portfolio values \\
Bank capital & Component and accounting equations & Capital ratios \\
Policy assessment & Structural model & Outcomes under alternative policies \\
\bottomrule
\end{tabularx}
\end{table}

The model runner receives a structured request instead of a prose recommendation. It executes a model only when the requested variables and outputs agree with a registered specification. If no specification matches the request, model execution stops for that request.

\subsection{Algorithm}

The algorithm assigns planning and interpretation to LLM agents, while retrieval, model execution, record checking, and numerical rendering follow explicit program rules. Each accepted request receives a unique analysis identifier, and every subsequent record is stored under that identifier. Opposing findings and rejection records remain under the same identifier, so no stage can silently modify an accepted source or replace a failed request, and an interrupted analysis can resume from the last completed stage without changing its inputs, identifiers, or model snapshot. The main stages are described below.

\paragraph{Planning and retrieval.}
The coordinating agent reads $q$ and identifies the target variables, outcomes, and admissible sources. Retrieval returns eligible passages and candidate relations under the restrictions in $\mathcal{C}_q$. The retrieval program constructs the graph queries and enumerates admissible directed paths, and the LLM ranks the candidates returned by that program.

\paragraph{Model request.}
The selected channels are translated into the variables, restrictions, and output fields required for calculation. Directional restrictions must be linked to recorded economic relations or to application rules fixed before execution, such as an accounting identity. The direction implied by a selected channel is composed by the program from the recorded relation signs, not narrated by the language model, and a channel whose composed direction contradicts the direction stated in its query cannot support a restriction. When the accepted channels support fewer restrictions than the model request schema requires, the affected analysis stops before any model call and records the reason. The request maps each variable to the controlled vocabulary and identifies the model specifications in $\mathfrak{M}_q$ that accept the required inputs, outputs, units, frequency, horizon, and scope.

\paragraph{Model execution and application calculation.}
The model runner executes every compatible specification and creates a separate \ModelRun{} for each calculation. The record identifies the model, input versions, parameters, environment, diagnostics, and output locations. Outputs from different models are not pooled into an ensemble, and no model is selected as the winner after the results are observed: each model retains its own tests and validation result. An application can then translate the model output into quantities such as portfolio values or bank capital. Each application result retains the identifier of the model run and the inputs used in the calculation.

\paragraph{Candidate selection.}
When several candidates satisfy the required restrictions and tests, the agent applies a selection criterion fixed before the corresponding evaluation results are inspected. A stress application can rank adverse scenarios, while an investment application can use a portfolio objective or risk constraint. Failed models and rejected candidates remain in the execution record instead of being replaced after their results are observed.

\paragraph{Report generation.}
The report generator receives the accepted evidence, selected channels, completed calculations, and application results. Numerical statements are rendered from registered model outputs after the corresponding tests have passed. Qualitative statements retain links to the relations and source locations used in the analysis. If a required relation, model input, or numerical test fails, the analysis stops at that stage and records the reason.

Our implementation uses specialized agents for coordination, evidence retrieval, model requests, execution, testing, consistency checking, and report generation. Their structured responses are validated before another stage can use them, and attempted calls remain in the execution record. Section~\ref{sec:multi-agent-execution} reports the saved executions used in the empirical analysis.

\subsection{Knowledge Base Construction}
\label{sec:financial-knowledge-graph}

The knowledge base is built before scenario generation so that document retrieval and graph search use the same registered source material. Each file receives a source identifier and hash, after which narrative documents and structured data are parsed into locations that can be retrieved without reparsing the original file during each analysis. Given the same file and parser version, the program produces stable identifiers for the extracted records, and a later download with a different hash receives a new source identifier; it does not replace the earlier file. When publication dates, observation dates, or data vintages are required by an application, these fields are stored with the corresponding locations rather than inferred from the date of the analysis.

Digitally generated PDF files are parsed by page. Text passages remain within a page so that retrieval can retain the page and quoted text, while structured tables retain their row and column labels with each numerical value. The parser also keeps baseline and adverse scenarios, jurisdictions, and data vintages in separate records when those distinctions affect the meaning of a value.

The relation database is constructed in two stages. The first stage identifies candidate text and maps each candidate to the predefined fields for an economic relation. The second stage validates the source location, variables, sign, and other permitted fields before the relation is written to the graph. For one graph construction condition, we also apply the decomposition principle in ATOM to split an eligible passage into short self-contained statements before relation extraction \citep{Lairgi2026atomadaptive}. The quotation and source location remain attached to each proposed record, so decomposition does not sever a claim from the passage from which it was obtained.

The extraction schema and the full relation representation in Equation~\ref{eq:sourcerelation} are kept distinct. The recorded extraction runs contain relation verbs and optional scope fields, while the full specification also distinguishes the evidence class and unknown scope. Missing information remains missing rather than being interpreted as universal applicability. When several sources support the same signed relation, their quotations remain separate, and conflicting signs remain associated with their own periods, jurisdictions, and empirical settings.

Source eligibility is determined before retrieval. In the European application, this rule is used to reconstruct the information available at a historical cutoff and to separate generation material from later evaluation material. Section~\ref{sec:empirical-analysis} describes the source collection and the rules for dated data used in the empirical analysis.

\subsection{Acceptance Tests}

Acceptance tests determine whether the output of one stage can be used by the next. The design adapts the logic of TDD to an analytical setting: the required properties of an output are fixed before that output is observed. These tests do not prescribe the unknown economic result. They specify conditions such as source eligibility, connected paths, compatible model inputs, required diagnostics, correct units, and consistency between a reported number and the model output from which it is drawn.

A numerical conclusion is a sentence that uses a number to describe an economic effect, compare outcomes, rank entities, or support a decision. For numerical conclusion $k$, the prespecified test is
\begin{align}
 \tau_k=\left(v_k,u_k,h_k,a_k,m_k,\rho_k,\epsilon_k\right),
 \label{eq:outputtest}
\end{align}
where $v_k$ is the output variable, $u_k$ is the unit, $h_k$ is the horizon, $a_k$ is the required transformation or aggregation, $m_k$ is the model specification, $\rho_k$ is the rounding rule, and $\epsilon_k$ is the permitted numerical tolerance. Required diagnostics, input versions, and application identifiers are stored with the same test.

Suppose that a candidate report uses value $z_k$ and that the linked model run returns $\widetilde{z}_{r,k}$, where subscript $r$ identifies fields stored with that run. The numerical test passes only if
\begin{align}
 v_k=v_{r,k},\quad u_k=u_{r,k},\quad h_k=h_{r,k},\quad a_k=a_{r,k},\quad
 \left|z_k-\operatorname{Round}_{\rho_k}\left(\widetilde{z}_{r,k}\right)\right|\leq\epsilon_k,
 \label{eq:outputtestpass}
\end{align}
and the model identifier, input versions, and required diagnostics match the prespecified test. If one sentence contains several numerical claims, each claim must pass its own test.

The recorded report interface uses \verb|{{NUM:claim_id}}| tokens for registered numerical values. The renderer resolves a token only after checking the model run identifier, file hash, source table location, unit, and stored value. The empirical evaluation also tests quantities written outside the registered fields, because successful token resolution alone cannot show that all other numerical statements are valid. Years and other temporal information are therefore represented separately from analytical quantities.

A failed acceptance test is recorded together with the stage at which it occurred. The agent cannot change the threshold, register a new quantitative function, or silently replace the failed record during the same run. These rules make stopping behavior part of the algorithm rather than an instruction that exists only in the prompt.

\section{Application: Stress Scenario Design}
\label{sec:stress-scenario-design}

We apply the general framework to macro-financial stress scenario design. This application is useful because the relevant risks can change quickly, while their consequences must be expressed as internally consistent paths for several economic and financial variables. The analysis therefore combines qualitative information about emerging risks with a quantitative model that produces a joint path and an application model that translates that path into financial outcomes.

The economic channels in Section~\ref{sec:agent} provide the link between these two parts of the analysis. An initiating disturbance can affect several outcomes through different mechanisms, and the graph can retain more than one channel for the same risk. The selected channels identify the variables and directional restrictions relevant to the scenario. A channel is accepted only when every relation has a verified source location, the variable and scope mappings are valid, and any apparent sign conflict is explained by the horizon, regime, or conditioning variables. In the European application, each channel query also registers the movement of its source variable and the expected movement of its target variable, so the program can compose the direction of a selected path from its recorded relation signs and refuse a path that contradicts the registered movement. Relations with opposing signs remain separate under their recorded horizons, regimes, and empirical settings. Scenario construction can therefore use the relation that matches the requested conditions without averaging findings that refer to different settings.

The macroeconomic model generates joint candidate paths subject to the admissible restrictions. A selection rule fixed before evaluation chooses adverse paths for the initiating risk, while an unconditional tail comparator selects adverse simulations without assigning them to a named risk. The selected macro-financial path can then be passed to a bank model that calculates capital and other balance sheet outcomes. The Appendix gives the formal definition of the stress scenario together with the execution map, severity measures, consistency diagnostic, and selection rule.

The European application uses information available by December 31, 2024 to construct scenarios for three broad risks: trade fragmentation, a property market correction, and market illiquidity. The dated source collection includes institutional financial stability assessments, research on economic transmission, structured scenario files, and risk surveys. Risk surveys are used to identify developments that were salient before the cutoff, while reports and research studies provide the relations used to construct candidate transmission channels. The later official stress scenario material is reserved for evaluation.

This application separates scenario generation from bank translation. The macroeconomic stage asks whether retrieved information can be converted into coherent restrictions and model simulations. The bank stage asks whether the resulting macro-financial paths can be mapped into outcomes for individual institutions using models that satisfy external validation criteria. This separation allows retrieval quality, scenario construction, and bank model performance to be evaluated as distinct parts of the same analysis.

\section{Empirical Analysis}
\label{sec:empirical-analysis}

The empirical analysis studies the European stress scenario application in Section~\ref{sec:stress-scenario-design}. It combines saved execution records with a set of rerun specifications that use consistent publication dates and numerical definitions. The saved runs establish which computational and agent operations were executed. The graph, scenario, and bank comparisons described below define the evaluation of economic performance under the specified data rules.

\subsection{Data and Information Date}

The complete parsed collection contains 51 files, including 41 PDF files and 10 spreadsheets. Parsing PDF files by page produces 2,744 page records and 4,188 text passages that do not cross page boundaries. Spreadsheet parsing produces 7,707 row records and 54,769 candidate numerical cells. Eligibility for generation is determined separately from these totals by the information date and the role assigned to each source.

The historical cutoff is December 31, 2024. Each numerical input identifies its observation period, the release date of the value actually used, and its data vintage; a quarter labeled by its starting date is not evidence that its value was public during that quarter. The rerun selects the last eligible release for each observation before applying transformations or aggregation, and no latest-vintage substitute is allowed when an eligible vintage cannot be obtained. The numerical model is estimated through the latest period for which every required series has an eligible observation under a common geographic definition. Later official scenario material and material reserved for development are excluded from the generation inputs used in that rerun. Generation inputs and official 2025 evaluation material are kept in separate directories, and hashes connect each stage to its actual inputs. Because the official 2025 material was inspected during development, it is a known retrospective benchmark rather than an unseen prospective outcome.

The three initiating risks use separate retrieval queries and five channel queries defined by source and target variables for each risk. Document retrieval ranks up to 20 eligible passages using reciprocal rank fusion of BM25 and rankings based on semantic similarity. Graph retrieval enumerates signed directed paths with no more than five relations and retains no more than five paths for LLM selection; a terminal return to the starting variable is allowed once for an explicitly specified feedback loop, while other repeated variables are rejected. Before restriction generation, the two evidence representations are serialized under the same limit of 12,000 bytes, with truncation only at record boundaries. Hybrid RAG allocates half of the limit to each representation and stops when their required directions conflict.

The recorded LLM operations use the OpenAI Responses API and the fixed model identifier \GPTFourOSnapshot{} \citep{OpenAI2024gpt4o}. The same requested snapshot performs relation extraction, path selection, restriction generation, and the agent tasks, and the identifier returned by the API is recorded with every response; a different or unavailable model stops the affected stage. Requests use temperature zero and strict JSON schemas, with server-side storage and automatic client retries disabled. Each completed request retains its payload, response, parsed output, token counts, hashes, and recorded cost. The same frozen prompts and inputs are used within a repeated experimental condition.

\subsection{Knowledge Graph and Retrieval Evaluation}

The G0 extraction condition identifies sentences that contain at least two variables from the controlled vocabulary and processes at most 600 candidates in batches of eight. A proposed relation is retained only when its source location is registered, its endpoints are two distinct variables from that vocabulary, and its quotation occurs in the cited text. These checks establish the structure of the record and its connection to a source passage. Whether the passage supports the attributed economic direction is evaluated separately.

The benchmark is constructed from the official 2025 scenario material and represents an economic mechanism either as one connected ordered path or as a set of connected paths. For a path, the target of each relation must equal the source of the next relation. Parallel effects are stored as separate paths rather than being scored as a two-step chain. The benchmark fixes source locations, endpoint variables, signs, and relation order in a machine-readable table before retrieval results are calculated; the extracted graph is compared with these external records and is never used as its own reference.

Let $g=(g_1,\ldots,g_m)$ be an accepted benchmark path and $p=(p_1,\ldots,p_n)$ a predicted path. Relation precision and recall use multiset overlap. Ordered scores use the length of the longest common subsequence, denoted by $\operatorname{LCS}(g,p)$. For a fixed benchmark variant, it holds that
\begin{align}
 \operatorname{OrderedRecall}(g,p)=\frac{\operatorname{LCS}(g,p)}{m}.
 \label{eq:orderedrecall}
\end{align}
The channel score is the best score among its accepted variants. Exact completion requires equality of the complete signed ordered sequence. Relation presence and endpoint reachability are reported separately from recovery of the ordered path.

\paragraph{Graph construction comparisons.}
The implementation distinguishes six graph construction conditions. G0 extracts relations from full sentences. G1 first decomposes each passage into short statements, and G2 adds a separate verification request for every G1 candidate. G3 expands the collection of official sources, G4 adds academic studies, and G5 ranks paths using verification status and support from independent sources. G0, G1, and G2 have saved execution records, while G3--G5 are specified comparisons for subsequent runs.

The recorded G0 and G1 runs differ in both decomposition and workload. G0 processes at most 600 sentences, while G1 processes 2,553 passages with one extraction request per passage. G2 then checks the 4,785 G1 records individually. The controlled decomposition comparison uses the same eligible passages, controlled vocabulary, response schema, and model snapshot in both conditions. The verification comparison applies G2 to exactly the G1 candidate set; because verification uses the same model snapshot as extraction, it provides a second assessment rather than an independent measurement of semantic correctness. Input coverage and computational expenditure are reported with the accuracy measures. Every downstream record stores the graph condition that produced it, so a construction result can be associated with a scenario result only when the downstream stages were executed with the same graph.

Evaluation first checks the benchmark path structure and then reports recovery of benchmark relations and connected path variants. Retrieval comparisons cover passage-constrained relation graphs, all enumerated paths, the five highest-ranked paths, deterministic ranking, and LLM selection, so a failure can be located before or after the selection step. Because the benchmark is not an exhaustive list of every correct economic relation, the fraction of extracted relations matching the benchmark is reported as the benchmark match fraction. Claims about the precision of the extracted relations require a separate assessment of whether each source passage supports the attributed direction. With three initiating risks, the main comparison reports results by risk rather than treating individual channels as independent replications. The saved G0 run retains 123 evidence records representing 81 distinct signed relations. G1 produces 4,785 relation records, and G2 retains 1,494 records representing 479 distinct signed relations after separate verification. These counts describe the output of the recorded construction stages; the controlled reruns described above are used for performance comparisons.

\paragraph{Source contribution.}
For a fixed graph and measure, let $M(G)$ denote the measure for graph $G$, and let $G_{-s}$ be the graph reconstructed without source document $s$. The contribution from removing document $s$ is
\begin{align}
 \Delta_s(M)=M(G)-M(G_{-s}).
 \label{eq:sourcecontribution}
\end{align}
Endpoint reachability, benchmark relation coverage, and recovery of complete ordered paths are calculated separately. The analysis also estimates average marginal contributions over 50,000 random source orderings with a fixed seed. This calculation assigns contribution according to the order in which sources are added to the graph and can therefore distinguish sources that are substitutable in the full graph.

\subsection{Scenario Generation and Quantitative Models}
\label{sec:graph-based-execution}

The scenario comparison includes a direct LLM condition as an experimental comparator together with three conditions that use retrieval. The direct condition asks the LLM for annual numerical paths without a quantitative model; numerical outputs from this condition are not accepted by the AI economist agent in Section~\ref{sec:agent}. Document RAG and GraphRAG translate their retrieved evidence into directional restrictions and select admissible paths from the same set of model simulations. Hybrid RAG combines the two forms of evidence under the common byte limit. Every downstream record stores the retrieval condition that produced it.

Restrictions concern the three-year mean of a variable relative to a specified reference unless the evidence identifies a particular timing. For each risk, the selection rule maximizes a fixed severity score over simulations satisfying the required restrictions. An unconditional comparator selects adverse simulations from the tail of the distribution without assigning them to the named risks. The weights and sensitivity analysis are stated in the Appendix.

The macroeconomic model is a quarterly vector autoregression with six variables, two lags, and 20,000 residual bootstrap simulations. Its estimation endpoint is determined by eligible data releases. Quarterly levels are retained so that annual and half-year quantities can be calculated under their stated definitions. The bank application specifies a dynamic quantile model with fixed effects and a linear model with fixed effects using the same accounting components. Estimation and evaluation features use the same transformations, periods, and geographic definitions.

External validation compares bank projections with official outcomes under the same hypothetical adverse scenario. A bank model must satisfy every specified validation threshold before its output for a generated scenario can support a conclusion for an individual institution. The comparison concerns agreement with official stress calculations under the common scenario rather than prediction of subsequently realized bank losses. The model specifications and thresholds are given in the Appendix.

\paragraph{Numerical and report checks.}
A numerical conclusion requires a successful model call and an accepted output. The recorded report interface uses registered numerical values, and the evaluation adds tests for quantities written in words, currency amounts, changes in basis points, malformed tokens, and incorrect placement of dates. The same acceptance rules are applied during generation and checking after execution. Further tests alter the stored numerical record, its source location, and its metadata to verify that the renderer rejects each inconsistency. Independent transformation calculations test the economic meaning of the value as well as equality with the stored output, so an error shared by the writer and the reader of a value cannot pass merely because both return the same number.

The graph evaluation reports retrieval coverage separately from whether the cited passages support the extracted relations. It also reports benchmark relation recovery, ordered path recovery, endpoint reachability, and source contribution under each graph condition. Scenario evaluation records the restrictions produced for each initiating risk and checks the transformations used to convert model simulations into annual quantities. The numerical evaluation distinguishes observed inputs, model outputs, and subsequent application calculations, while report evaluation checks source references and registered numerical statements against their model runs. The empirical comparison records the graph condition with every selected path, restriction, model run, and report. The Appendix gives the full definitions of these measures and tests.

\subsection{Multi-Agent Execution}
\label{sec:multi-agent-execution}
\label{sec:bounded-agency}

We use the saved multi-agent runs to examine whether the implementation follows the programmed stages and stopping rules. The program assigns structured tasks to seven agents for coordination, evidence retrieval, model requests, execution, testing, consistency checking, and report generation. The execution program controls the registered model calls and acceptance rules. Each attempted request remains in the record together with its response, status, and relevant identifiers.

\paragraph{Recorded runs.}
The system was executed under seven prompt configurations with the same three initiating risks. All configurations use the G0 graph constructed from full sentences and the associated path records. Run 1 stops at the first coordinator response because the prompt does not supply identifiers required by the audit interface. Run 2 reaches model execution and reporting but later encounters missing prefixes required for numerical tokens. Run 3 supplies the complete token syntax and completes the programs for all three risks in three repetitions, while two repetitions stop at report generation because a calendar year is placed inside a numerical token.

In Runs 1 through 3, the direction restrictions supplied to the model were assigned from the risk registration, with the selected path recorded as a reference. Runs 4 and 5 instead derive the restrictions from the channels selected by the evidence agent, under the composition rule in Section~\ref{sec:agent}: the program composes the direction of each selected path from its recorded relation signs, refuses a path whose composed direction contradicts the registered movement of its channel query, and writes a derivation record naming every assessed path, the derived restrictions with their supporting relations, and the digest of the registration it ran under. Under this rule, both recorded paths for market illiquidity compose against their registered movements, so that risk program stops before any model call in every repetition. This stop is a deterministic consequence of the recorded graph and is retained as a result: given the recorded paths, the attainable maximum for these configurations is completion of the two remaining risk programs.

In Run 4, the remaining programs also fail to complete. The model request agent writes a digit into its free-text rationale when restating the assigned horizon, the numeric contract refuses the response as a contract violation, and the repetition ends before trade fragmentation is reached; at temperature zero the same response recurs in every repetition. Run 5 restates the prompt instruction that the rationale must be written in words alone. Three of its five repetitions complete both derivable risk programs and their reports. One repetition stops when the evidence agent cites a relation identifier that differs from a recorded identifier by one character, and one stops at report generation when a directional sentence cites no accepted relation.

The completed Run 5 reports satisfy every reference rule and still read as a table: parallel one-clause sentences, one value each. Run 6 therefore adds the instruction that the report be written as connected prose. The instruction collides with the registered output budget: in every repetition but one, the longer report exceeds the registered limit of 1,024 output tokens, the truncated response cannot be parsed, and the affected program stops with an unusable response. Run 7 raises the registered output limit to 2,048 tokens and changes no prompt. Every Run 7 repetition completes the property correction program and two also complete trade fragmentation; the other three stop at report generation when the report cites relation identifiers inside numerical tokens, which the numeric contract refuses as unknown claims. The generated reports are reproduced in Appendix~\ref{app:generated-reports}.

\begin{table}[!htbp]
\centering
\caption{Recorded execution by prompt configuration. Completed and stopped counts refer to individual risk programs. Each configuration schedules five repetitions of three risks. From Run 4 onward, the stopped counts include the market illiquidity program, which stops before its model call in every repetition because its restriction derivation refuses both recorded paths.}
\label{tab:agent-runs}
\begin{tabular}{lrrrrr}
\toprule
Configuration & Calls & Cost (USD) & Completed & Stopped & Not started \\
\midrule
Run 1 & 5 & 0.0164 & 0 & 5 & 10 \\
Run 2 & 63 & 0.2908 & 4 & 5 & 6 \\
Run 3 & 77 & 0.3555 & 9 & 2 & 4 \\
Run 4 & 25 & 0.1442 & 0 & 10 & 5 \\
Run 5 & 68 & 0.3801 & 7 & 7 & 1 \\
Run 6 & 53 & 0.3129 & 1 & 10 & 4 \\
Run 7 & 81 & 0.4883 & 7 & 8 & 0 \\
\bottomrule
\end{tabular}
\end{table}

Across the seven configurations, 372 LLM calls incur a recorded cost of 1.9882 USD. The saved records include failed calls as well as successful calls, so the completion counts can be read together with the rules that stopped each run.

\begin{figure}[!htbp]
\centering
\includegraphics[width=0.98\linewidth]{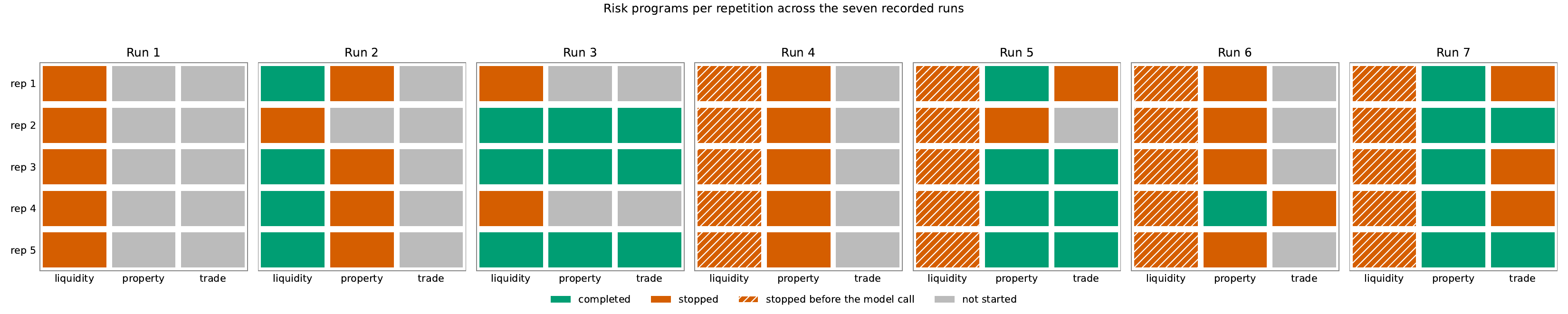}
\caption{State of each risk program by repetition and configuration. No repetition in Runs 1 or 2 completes all three risks, and in Run 3 three of five repetitions do. From Run 4 onward, the market illiquidity program stops before its model call in every repetition because the restriction derivation refuses both recorded paths, so the attainable maximum is completion of the two remaining programs; three Run 5 repetitions and two Run 7 repetitions reach it, and every Run 7 repetition completes the property correction program.}
\label{fig:agent-progression}
\end{figure}

\paragraph{Record checks and agreement.}
The saved checks performed after execution verify hashes of step records, references to evaluation material, digests of model outputs, and registered numerical values read from the referenced tables. The checks verify the integrity of the saved records, including records for runs that stopped before completing the full analysis. Table~\ref{tab:agent-audit} reports the objects inspected by configuration.

\begin{table}[!htbp]
\centering
\caption{Objects inspected by the saved checks performed after execution. Claims are registered numerical values read from source tables, whether or not a final report was completed. From Run 4 onward, the integrity records include the restriction derivation records, which the checks recompute from the recorded channel selections and compare with the stored records field by field.}
\label{tab:agent-audit}
\begin{tabular}{lrrrr}
\toprule
Configuration & Step records & Model runs & Integrity records & Claims read \\
\midrule
Run 1 & 5 & 0 & 0 & 0 \\
Run 2 & 63 & 9 & 27 & 162 \\
Run 3 & 77 & 11 & 33 & 198 \\
Run 4 & 25 & 0 & 10 & 0 \\
Run 5 & 68 & 8 & 37 & 144 \\
Run 6 & 53 & 6 & 29 & 108 \\
Run 7 & 81 & 10 & 45 & 180 \\
\bottomrule
\end{tabular}
\end{table}

Agreement is measured over the union of risks reached by each pair of repetitions, with an unreached risk counted as disagreement. In Run 3, agreement on the selected paths is 0.571 across all ten pairs and the Jaccard similarity of warning subjects is 0.548; the three repetitions that complete every risk select the same paths and warning subjects. In Runs 4 and 5, an evidence selection names one declared path for each selected channel, and two selections agree only when they name the same set. Run 4's five repetitions return identical selections and stop at the same rule, so path agreement equals one. In Run 5, path agreement is 0.793 and the Jaccard similarity of warning subjects is 0.731; the two repetitions that stop differ from their siblings through single responses rather than through different selected evidence. In Run 6, the early report stops leave trade fragmentation unreached in four repetitions, and path agreement is 0.833 with warning similarity 0.769. In Run 7, both measures equal one: every repetition selects the same channel sets and warns about the same subjects. These statistics describe agreement under the same candidate evidence and numerical settings.

\begin{figure}[!htbp]
\centering
\includegraphics[width=0.98\linewidth]{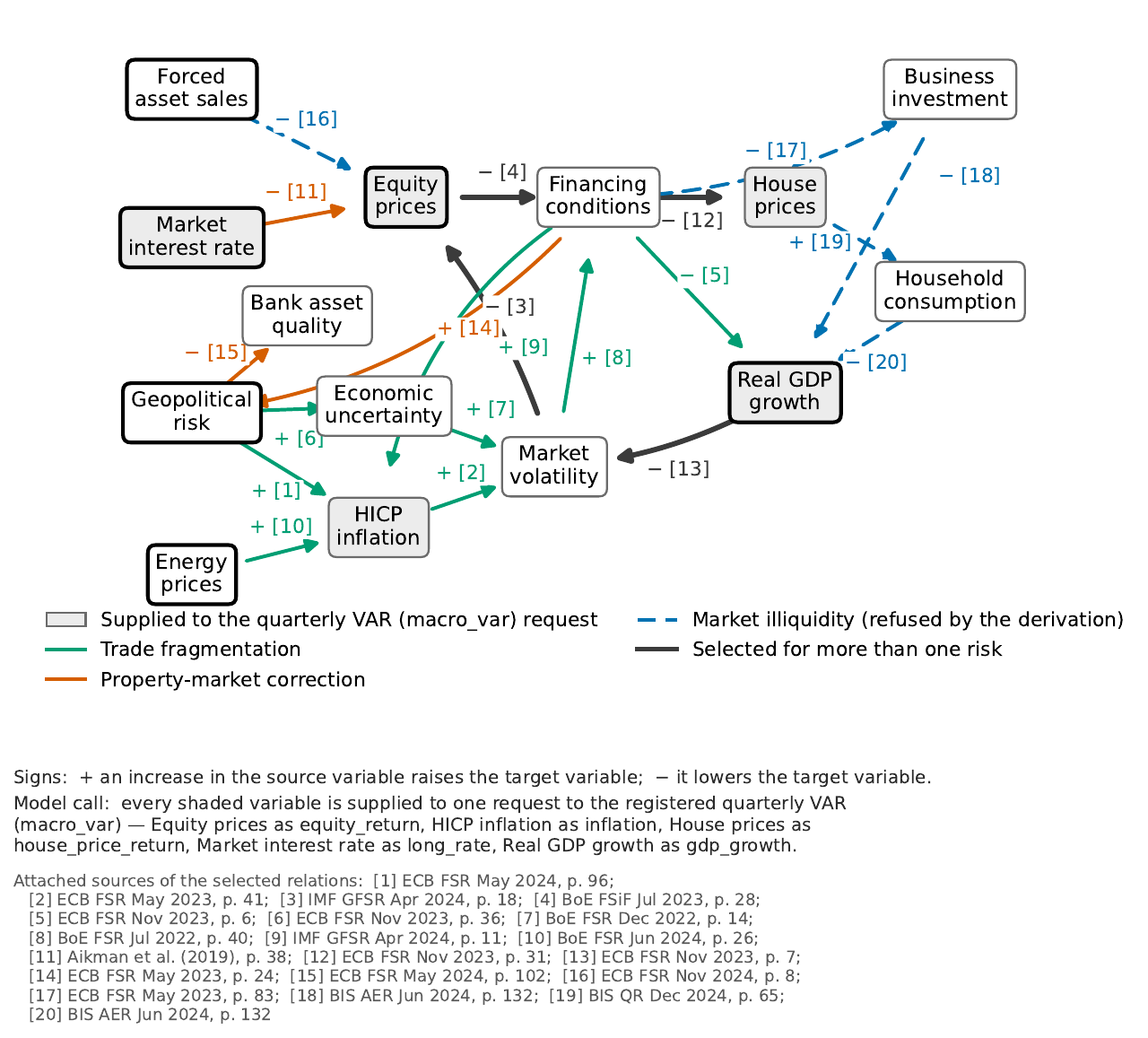}
\caption{Signed transmission paths selected in the recorded Run 7 repetitions. Boxes are variables from the controlled vocabulary, and each arrow is one signed relation from a selected G0 path constructed from full sentences. A plus sign states that an increase in the source variable raises the target variable, and a minus sign that it lowers the target variable: relation [4], for example, states that lower equity prices tighten financing conditions. Shaded boxes mark the variables whose derived restrictions are supplied to the single model request, issued to the registered quarterly vector autoregression (\texttt{macro\_var}); the key at the foot of the figure names the model variable each shaded box maps to. The bracketed key identifies the source document and page listed at the foot of the figure. Boxes with thick borders mark the source variables of the channel queries. Dashed arrows belong to the two market illiquidity paths, which the restriction derivation refused, so no restriction rests on them. The source key identifies where each relation was extracted; the graph experiment evaluates whether the passage supports the displayed direction.}
\label{fig:selected-paths}
\end{figure}

Figure~\ref{fig:selected-paths} displays the paths selected in the recorded Run 7 repetitions, together with the source attached to each relation; the five repetitions select identically, and the selection matches the complete Run 5 repetitions. Every declared channel path is selected for every risk, and the two market illiquidity paths are drawn dashed because the restriction derivation refused them. Manual inspection shows that three displayed G0 relations are not supported by the attached text. The relation from financing conditions to business investment is attached to a quotation about losses of investment funds. The relation from equity prices to financing conditions is attached to a passage that does not establish that direction, and the relation from market volatility to equity prices is attached to an incomplete extracted sentence. These errors are included in the evaluation of whether the cited passages support the extracted relations.

The report interface in Runs 1 through 3 audited registered numerical claims without assigning a relation identifier to every qualitative report sentence. In Runs 4 and 5, every model restriction identifies its supporting relations or the registered rule under which it is imposed, and a directional report sentence must cite an accepted relation; this rule is what stops one Run 5 repetition at report generation. The checks performed after execution recompute the derivation record of every recorded channel selection from the registered artefacts and accept all five configurations.

\section{Conclusion}
\label{sec:conclusion}

We propose an AI economist agent that combines evidence retrieval, economic transmission channels, quantitative models, and report generation in one framework. The central idea is to use LLM agents for tasks that require flexible interpretation and search while assigning numerical conclusions to registered models with explicit inputs, outputs, and acceptance tests. This structure allows the analysis to adapt to new questions without allowing the language model to replace the calculation on which a quantitative conclusion depends.

The European application shows how the framework can be used for stress scenario design and bank capital analysis. The empirical analysis examines construction and retrieval of economic relations, generation of macro-financial scenarios, validation of numerical reporting, and execution of the multi-agent program. The saved runs show that the model calls, the derivation of restrictions from selected channels, and the stopping rules can be enforced, while the graph evaluation identifies where relation extraction still fails to recover the intended economic direction.

The framework is intended as a general method for economic analysis in settings where the relevant evidence and the appropriate quantitative model vary with the question. By keeping source material, economic relations, model specifications, calculations, and reported values as distinct records, the AI economist agent provides a way to combine flexible scenario construction with reproducible quantitative analysis.

\section*{Data and Code Availability}

The replication materials contain source manifests, code, fixed configurations, saved API records, numerical outputs, notebooks, and tests. Sources are identified by location and hash, model run records retain their requests and input versions, and agent records include attempted calls and termination reasons. The saved API executions reported in Section~\ref{sec:multi-agent-execution} are included with the replication materials. The numerical comparisons that use the release-date rules in Section~\ref{sec:empirical-analysis} will be reported after the corresponding estimation and reruns.

\clearpage
\appendix

\section{Extended Related Work}

The proposed design draws on research about economic analysis with LLMs and methods that combine retrieval with quantitative modeling. We discuss these general elements before reviewing the stress-testing literature that supports the principal empirical application.

\paragraph{LLM agents and external computation.}
Generative AI can assist with research tasks such as reviewing literature, processing data, and writing code \citep{Korinek2023generativeai}. Systems with tool use and persistent state can connect these tasks \citep{Korinek2025aiagents}, although each step creates a distinct source of error. Retrieval may omit contrary evidence or graph search may select the wrong relation; later stages can accept incompatible units or describe a calculation that did not occur.

The algorithm assigns checks to the interfaces at which these errors can reach the analysis. Before an economic relation is added to the graph, the extraction program verifies its source location and required fields, while the LLM fills predefined query fields rather than writing unrestricted database code. The model runner accepts only functions identified in the supplied specifications and stores the hashes needed to reproduce each calculation. Finally, the report checker compares qualitative statements with their cited sources and applies tests fixed before model execution to every numerical conclusion. In the financial application, these checks also connect analysis with LLMs to the principles for risk data aggregation and reporting used in financial supervision \citep{BCBS2013principlesfor}.

\paragraph{Test-driven development and analytical requirements.}
Test-driven development places executable tests before the implementation they guide \citep{Beck2002testdriven}. We use this idea to specify the interfaces between evidence processing and numerical calculation. A unit test may have a known numerical answer, whereas a test on a scenario model specifies required units, transformations, and identities without prescribing the unknown scenario outcome. The algorithm uses those tests to govern acceptance during execution. The run records show application of the encoded rules, not a complete chronological record of the development of every software component.

\paragraph{Retrieval and knowledge graphs.}
Retrieval-augmented generation combines a language model with external information retrieval \citep{Lewis2020retrievalaugmented}. Although standard document RAG can retrieve passages that mention an economic condition or entity, a question about how that condition affects an outcome often requires evidence from several sources. One source may describe the initial price response, another the change in policy or financing conditions, and a third the later economic or financial outcome. When no single passage contains the complete channel, document retrieval leaves the LLM to infer how the separate findings fit together.

Retrieval over a graph represents relevant entities and relations before generation. The GraphRAG method in \citet{Edge2025fromlocal} uses graph structure and hierarchical summaries for questions whose evidence is distributed across a corpus. HippoRAG links passages through a knowledge graph to support multi-hop retrieval \citep{Gutierrez2024hipporagneurobiologically}, whereas LightRAG combines local and global graph search with efficient indexing \citep{Guo2025lightragsimple}. The evaluation of economic channels requires a specified scope and a connected relation sequence. Our comparison fixes path length and source eligibility together with the query scope and retrieval limits for each method before evaluating recovered channels against reviewed benchmark paths.

ATOM decomposes source text into minimal self-contained facts before extracting temporal five-tuples, separates observation dates from validity periods, and merges the resulting temporal graphs in parallel \citep{Lairgi2026atomadaptive}. Decomposition supplies candidate facts, whose support still has to be assessed against the source quotation. We adopt the decomposition principle for the extended graph construction, and retain a quotation with each proposed relation. The specification additionally distinguishes the evidence class from the relation verb and requires explicit treatment of unknown scope. The separate verification request assesses the source support for a candidate before the candidate is written to the graph.

Macroeconomic knowledge graphs provide a domain precedent: \citet{Yang2020knowledgegraph} represents relations among macroeconomic variables and alternative data for forecasting applications. Related work also clarifies the roles of information generated or retrieved by an LLM. \citet{Kameyama2026causalityelicitation} derives candidate causal graphs from documents generated repeatedly by an LLM and treats those graphs as inspectable hypotheses rather than verified real-world relations. \citet{Kato2026vectorsearch} formulates action selection with RAG as policy learning and connects vector search tailored to the action to nearest-neighbor matching. In our setting, graph relations come from dated external sources and constrain requests to financial models. A retrieved relation arrives as a candidate with a source reference, not as an action or a settled causal claim. A separate verification request checks whether the cited passage supports the attributed direction.

Our algorithm connects requirements for sources to requirements for numerical execution. It retains the source and available scope information for each relation, while calculated values are distinguished from observations and linked to completed model runs. Research on citation generation and RAG evaluation is also relevant \citep{Gao2023enablinglarge,Es2024ragasautomated,SaadFalcon2024aresan}. For economic and financial analysis, the cited source must support the relation under the stated scope, and every numerical conclusion must match a recorded output.

\paragraph{Quantitative models for economic and financial analysis.}
A qualitative analysis becomes quantitative when it imposes restrictions or conditions on a numerical model. \citet{Waggoner1999conditionalforecasts} derive conditional forecasts for dynamic multivariate models, while \citet{Banbura2015conditioalforecasts} show how restrictions on selected variables can be propagated through a large vector autoregression. Cross-country models can represent foreign influences through trade or financial weights \citep{Pesaran2004modelingregional,Chudik2016theoryand}, whereas local projections are useful when the object is a horizon-specific response to a well-defined shock \citep{Jorda2005estimationand}. Structural sign and zero restrictions permit shock interpretation without a complete recursive ordering \citep{Arias2018inferencebased}, and narrative restrictions can add information from dated historical episodes \citep{AntolinDiaz2018narrativesign}. These methods can be supplied through the quantitative model specification developed below, but none defines the general agent.

The appropriate diagnostic depends on the analytical task. In a conditional macroeconomic projection, the central concern may be whether the variables form a joint scenario consistent with the selected multivariate model. An investment analysis may instead require valuation and exposure models, while a corporate analysis can connect sector conditions to financial statements. Growth-at-risk models illustrate how financial conditions affect the lower tail of future activity \citep{Adrian2019vulnerablegrowth}, and geopolitical-risk measures provide dated indicators for one source of global shocks \citep{Caldara2022measuringgeopolitical}. The model specifications described below accommodate these differences by stating the inputs and diagnostics required by each application.

\paragraph{Stress testing and scenario design.}
Bank stress testing combines an adverse scenario with models of the balance sheet and capital. Surveys of supervisory practice document differences in granularity and feedback effects, as well as in the division between top-down and bottom-up calculations \citep{Foglia2009stresstesting,BCBS2017supervisoryand,Baudino2018stresstesting}. Despite these differences, the scenario is the common input that connects a macroeconomic disturbance to bank losses and capital. The Basel Committee accordingly emphasizes well-defined objectives and governance supported by reliable data, models, and communication \citep{BCBS2018stresstesting}. Within this broader exercise, top-down models translate scenario values into outcomes for individual banks. For example, the dynamic panel quantile regressions in \citet{Covas2014stresstestingus} allow losses and revenues to respond nonlinearly during stress, whereas the Capital and Loss Assessment under Stress Scenarios model in \citet{Hirtle2016assesingfinancial} uses public data to project bank income and capital. The European method in \citet{Henry2013amacro} likewise combines macroeconomic scenarios with bank satellite models and contagion channels. Together, these studies motivate the empirical separation between the model that produces macro-financial values and the equations that translate those values into bank outcomes.

Stress tests can also reveal information about financial institutions. \citet{Flannery2017evaluatingthe} study the market information in Federal Reserve stress tests, while \citet{Philippon2017backtestingeuropean} compare projected and realized outcomes in European exercises. The analyses in \citet{Acharya2014testingmacroprudential} and \citet{Goel2021limitsof} show that regulatory measurement and institutional constraints affect the conclusions drawn from stress tests. Accordingly, our empirical evaluation examines losses and vulnerability rankings rather than relying on narrative quality alone.

The literature on scenario design provides a basis for selecting more than one adverse scenario. \citet{Parlatore2025designingstress} formulate a stress scenario as an information choice that reveals unknown exposures and supports later intervention. \citet{Aikman2024stresstesing} search across several macro-financial scenarios and identify those associated with adverse outcomes for the banking system and individual institutions. These studies motivate evaluating more than one adverse case. Our specified implementation uses severity selection for each risk and a separate unconditional clustering rule; it does not estimate the information objective studied by Parlatore and Philippon. Other stress exercises show that the relevant relation can depend on its empirical scope \citep{Aikman2019systemwidestress,Acharya2023climatestress,Baudino2024liquiditystress}, so the graph retains that information.

\section{Retrieval and Graph Search Implementation}

The main text summarizes retrieval and graph search. This section gives the corresponding parameters, path ranking score, and records stored by each operation of the algorithm.

\subsection{Source Retrieval}

After applying the source eligibility rules in the analysis request, the retrieval step uses BM25 and latent semantic retrieval over the same eligible source locations. The semantic representation applies truncated singular-value decomposition to term frequency and inverse document frequency features. Reciprocal rank fusion combines the lexical and semantic rankings. In the empirical application, each search returns 50 candidates, after which no more than 20 source passages are passed to subsequent steps.

Document RAG and GraphRAG use the same source collection and information date rule. Document RAG is limited to 20 passages, while GraphRAG returns no more than five paths and no path may contain more than five relations. Thus, graph retrieval cannot improve its result by drawing on a later source collection or by returning an unbounded path.

Excluded sources remain associated with the analysis identifier together with the reason for exclusion. A source can be excluded because it falls outside the permitted information set, concerns an incompatible jurisdiction or sector, belongs to a disallowed category, or fails the check of its exact source location after normalization. Keeping these exclusions permits error analysis of the eligibility rules.

\subsection{Graph Search}

The LLM does not write queries for graph search. The retrieval program receives the initiating condition and target outcome from the generation registry during generation and from the separate benchmark during evaluation, enumerates directed paths in the accepted relation table, and enforces the limits on path length and eligible sources.

The graph baseline chooses the path with the highest score without asking the LLM to compare the returned sequences of relations. GraphRAG presents no more than five admissible paths to the fixed GPT-4o snapshot, which selects one path using only the attached relation evidence. If the relation table contains no admissible path, both graph conditions return no path for that query.

Opposing findings remain separate. For example, one study may find that tighter policy lowers corporate investment, whereas another may find a weak response among firms with large cash holdings and long debt maturities. The graph stores both relations with their horizons and conditioning variables, which allows the model request to state where each relation applies instead of using an average sign that neither source supports.

\subsection{Path Ranking and Stored Results}

For an admissible path $C$ containing $L$ relations, the implemented ranking score is
\begin{align}
 S(C)=\frac{1}{L}\sum_{\ell=1}^{L}q_{\ell}+0.05n_{\mathrm{doc}}(C)-0.03(L-1),
\label{eq:channelscore}
\end{align}
where $q_{\ell}$ is the extraction confidence for relation $\ell$ and $n_{\mathrm{doc}}(C)$ is the number of distinct documents attached to the path's relations. The first term uses confidence assigned by the extractor rather than independently measured source support, the second gives a small reward to paths whose relations are attached to several documents, and the last discourages unnecessary intermediate relations. For the planned comparison, paths are grouped only when both the ordered variables and relation signs agree. A different sign pattern remains a distinct candidate even when the variable sequence is identical. Equation~\ref{eq:channelscore} ranks the available evidence for a specified query but does not determine whether an economic relation is true.

Table~\ref{tab:execution-groups} states the record stored after each operation of the algorithm.

\begin{table}[!htbp]
\centering
\caption{Information stored during an analysis.}
\label{tab:execution-groups}
\begin{tabularx}{\linewidth}{p{0.35\linewidth}Y}
\toprule
Operation & Stored result \\
\midrule
Request checks & Accepted request or rejection reason \\
Source retrieval & Source locations and exclusion reasons \\
Graph search & Ranked connected paths \\
Model request & Inputs, restrictions, and output tests \\
Model execution & Run identifier, outputs, and diagnostics \\
Application calculation & Results for the application \\
Candidate selection & Selected identifiers and score values \\
Output tests & Test results and rejected claims \\
Report generation & Checked report or termination record \\
\bottomrule
\end{tabularx}
\end{table}

\section{Stress Scenario Formalism}

This section defines the objects passed from qualitative scenario construction to quantitative execution. It retains the distinction in the main text between a directional restriction carrying attached evidence and a numerical value calculated under that restriction.

\subsection{Definition of a Stress Scenario}

We represent a qualitative scenario $s$ as
\begin{align}
 s=\left(d_s,\mathcal{C}_s,\mathcal{V}_s,\Sigma_s,\mathcal{T}_s,\mathcal{J}_s,\mathcal{K}_s,H_s,\mathcal{E}_s,\mathcal{E}_s^{-}\right),
 \label{eq:scenario}
\end{align}
where $d_s$ identifies the initiating disturbance and $\mathcal{C}_s$ contains its accepted economic channels. The set $\mathcal{V}_s$ contains variables requiring qualitative or numerical treatment. Directional restrictions are stored in $\Sigma_s$, and $\mathcal{T}_s$ contains timing restrictions stated in the attached evidence. The sets $\mathcal{J}_s$ and $\mathcal{K}_s$ identify jurisdictions and sectors, while $H_s$ is the horizon. Supporting and opposing evidence are stored separately in $\mathcal{E}_s$ and $\mathcal{E}_s^{-}$.

The language model completes these fields without assigning the calculated magnitudes. For example, evidence may support a decline in output and a rise in unemployment following trade fragmentation. The direction can then be supplied to a model request, whereas the percentage changes must come from the subsequent numerical calculation. Observations used to describe initial conditions remain labeled as observed inputs.

A candidate proceeds to calculation only when every required relation belongs to an accepted connected channel. Parallel mechanisms use separate paths. An intermediate relation lacking an accepted relation record ends the affected analysis with a record of the missing connection, rather than being supplied implicitly by the language model.

\subsection{Inputs to the Quantitative Model}

Let $\mathfrak{M}$ denote the model specifications available to the application. Each $M\in\mathfrak{M}$ defines its input and output requirements together with an implementation and a data or parameter snapshot. The request for scenario $s$ is
\begin{align}
 r_s=\left(s,I_s,O_s,H_s,\nu_s\right),
 \label{eq:modelrequest}
\end{align}
where $I_s$ and $O_s$ are the requested inputs and outputs, and $\nu_s$ identifies their data and parameter versions.

The economic channels determine the required variables. A rate shock transmitted through property values and credit losses requires different application inputs from a bond valuation calculation. The request records whether a restriction follows from retrieved evidence, an accounting identity, or an analytical assumption fixed before evaluation. It can therefore be checked before any model is run.

The execution program compares the request with every specification in $\mathfrak{M}$ and executes all compatible models separately. If none accepts the required variables, scope, and horizon, the analysis ends without a numerical scenario.

\subsection{Numerical Calculation}

Let $D$ denote the numerical input data and $\mathcal{R}_s$ the restrictions implied by the scenario, including any accepted timing restriction. A compatible model returns
\begin{align}
 \left(z_{s,M},U_{s,M},\Delta_{s,M}\right)=\operatorname{Execute}\left(M,r_s,D,\mathcal{R}_s\right),
 \label{eq:execute}
\end{align}
where $z_{s,M}$ is the numerical scenario, $U_{s,M}$ contains uncertainty information when provided, and $\Delta_{s,M}$ contains diagnostics. The result record retains the request and model specification together with the data and implementation hashes.

A probabilistic model with conditional distribution $p_M$ can produce a draw satisfying
\begin{align}
 z_{s,M}\sim p_M\left(z\mid D,\mathcal{R}_s\right).
 \label{eq:probabilisticscenario}
\end{align}
A deterministic model instead returns $z_{s,M}=g_M(D,\mathcal{R}_s)$ for its specified execution function $g_M$. The stored output distinguishes a point projection from a predictive interval or a collection of admissible trajectories. Restrictions on a reduced-form trajectory do not, by themselves, identify a structural economic shock.

Global vector autoregressions and local projections can represent transmission at different levels of aggregation. Structural models can impose behavioral equations, while accounting models translate losses into capital. The algorithm admits these models through their specifications without depending on one particular class.

\subsection{Severity of Individual Variables}

Severity measures how far a variable moves into a specified adverse tail. For variable $v$, horizon $h$, and model $M$, let $F_{M,v,h}$ denote the cumulative distribution function selected as the reference. When an increase is adverse, the severity percentile is
\begin{align}
 \pi_{s,M,v,h}=F_{M,v,h}\left(z_{s,M,v,h}\right).
 \label{eq:severitypercentile}
\end{align}
When a decline is adverse, it is $1-F_{M,v,h}(z_{s,M,v,h})$. The reference distribution and adverse direction are recorded for each variable. Marginal percentiles are reported separately from any criterion used to select an entire path.

\subsection{Consistency of the Joint Scenario}

Joint consistency concerns whether the variable values can occur together under the chosen model and restrictions. A simulation model can report the share of draws satisfying all restrictions, while a structural model can report equilibrium residuals. For a diagnostic function $J_M$, we write
\begin{align}
 J_M(s)=J_M\left(z_{s,M},D,\mathcal{R}_s\right).
 \label{eq:jointdiagnostic}
\end{align}
Its definition is fixed in the model specification. A restriction-acceptance frequency describes the simulation distribution under the model; it is not a probability assigned to the initiating narrative by the language model.

Every numerical path must also satisfy its stated unit, range, and metadata checks. Required model diagnostics, such as stability or accounting identities, are checked before the output is passed to an application calculation. A rejected path is retained in the execution record with its failed condition.

\subsection{Selection of Several Scenarios}

The implemented selection separates scenarios for each initiating risk from unconditional adverse paths. Let $\mathcal{K}_{\mathrm{risk}}$ be the set of three initiating risks, and let $\mathcal{A}_k$ contain the simulated paths satisfying every requirement for risk $k$. The severity score $S_{s,k}$ for risk $k$ is defined in Equation~\ref{eq:restrictionseverity}. The selection rule is
\begin{align}
 s_k^{*}\in\arg\max_{s\in\mathcal{A}_k}S_{s,k},\qquad
 \mathcal{S}^{*}=\{s_k^{*}:k\in\mathcal{K}_{\mathrm{risk}}\}.
 \label{eq:scenarioset}
\end{align}
The algorithm resolves an exact tie by the smallest stored simulation identifier. When $\mathcal{A}_k$ is empty, the corresponding risk program stops. Different risks may select the same numerical path, so the number of distinct selected paths is reported as an outcome rather than forced to equal three.

The unconditional comparator retains the most adverse decile under a fixed standardized severity score, partitions its feature vectors into three groups using $K$-means, and selects the most severe member of each group. Those paths retain neutral identifiers. This clustering rule and the severity maximization for each risk are the selection methods evaluated here. An application that chooses scenarios according to information for individual institutions would require its own explicit objective and validation design.

\section{Scenario Construction and Quantitative Models}

This section specifies the numerical evaluation design. It defines compatible transformations before estimation and states the rules for scenario selection and bank validation. Results from execution of this design based on dated data are not reported.

\subsection{Scenario Construction and Evaluation}

The direct condition requests annual values for 2025--2027 from the LLM without retrieved evidence or a numerical model. It is an experimental comparator, not an accepted mode of the agent. Document RAG and GraphRAG convert their evidence into directional restrictions and select from a common simulation set. Hybrid RAG combines non-conflicting restrictions under the same total limit on evidence size in bytes. An unconditional tail condition selects three paths with neutral identifiers and does not assign them to the named initiating risks.

Each restriction concerns the three-year mean of a specified annual variable. GDP growth and changes in asset prices use zero as their reference. Inflation, unemployment, and the long-term interest rate use the corresponding last eligible annual observation under the same definition as the projected value. A reference is neither taken from an unpublished quarter nor supplied by the LLM. Timing for a specific year is imposed only when the source supports it.

\paragraph{Severity by risk.}
Let $N=20{,}000$ be the number of simulated trajectories, and let $\bar{x}_{s,j}$ be the three-year mean of annual variable $j$ for simulation $s\in\{1,\ldots,N\}$. The values $\mu_j$ and $\sigma_j$ are its mean and standard deviation over that common simulation set. For risk $k$, there are $R_k$ restrictions. Restriction $r$ refers to variable $j(r)$, has priority $q_r\in\{1,2,3\}$, and has direction $d_r=1$ for an increase or $d_r=-1$ for a decrease. The primary severity score is
\begin{align}
 S_{s,k}=\frac{\sum_{r=1}^{R_k}(4-q_r)d_r(\bar{x}_{s,j(r)}-\mu_{j(r)})/\sigma_{j(r)}}{\sum_{r=1}^{R_k}(4-q_r)}.
 \label{eq:restrictionseverity}
\end{align}
The rule applies only when $R_k\geq 1$ and every required $\sigma_{j(r)}$ is positive. It selects from paths that already satisfy all restrictions. Equal weights and weights $1/q_r$ provide the two specified sensitivity comparisons. The mapping $4-q_r$ is an analytical choice and is frozen before this execution; it is not presented as a historical preregistration. In the configurations that derive restrictions from selected channels, the priority $q_r$ is assigned by a registered rule: a restriction restating the registered movement of a channel's initiating variable has priority one, a restriction carried to a channel's registered target variable has priority two, and a restriction read from an intermediate variable has priority three.

The risk is the descriptive unit of the comparison. The analysis reports each risk's directional violations and the difference between methods, together with the number of nonzero differences and ties. A distance from the official European scenario is computed only for quantities sharing the same definition and unit. It measures resemblance to a common hypothetical path, not forecast accuracy for each risk.

\subsection{Numerical Definitions and Temporal Aggregation}

The input registry preserves positive levels for GDP and price indexes in addition to the transformed variables used for estimation. It records whether a level is a period average or an end-of-period observation. A transformation identifier follows the value through the model request, output record, and report.

\paragraph{Log changes and percentage changes.}
For a positive quarterly level $P_t$, define $\ell_t=100\log(P_t/P_{t-1})$. The ordinary percentage change between quarters $a$ and $b$ is $100(\exp(\sum_{t=a+1}^{b}\ell_t/100)-1)$. Summing log changes is appropriate for the log change over that interval, but the sum is not labeled as an ordinary percentage return. The same rule applies when combining two quarters into a half-year change.

\paragraph{GDP and consumer prices.}
Let $Y_{y,q}$ be the real GDP level for quarter $q$ of calendar year $y$. The annual GDP growth rate is $100((\sum_{q=1}^{4}Y_{y,q})/(\sum_{q=1}^{4}Y_{y-1,q})-1)$. Half-year GDP growth compares the sum of the two quarters in a half-year with the corresponding sum in the preceding half-year. These definitions differ from the cumulative change between two quarterly endpoints.

Let $H_t$ be the quarterly average harmonised index of consumer prices (HICP). The internal inflation variable is the year-on-year log change $\pi_t^{\log}=100\log(H_t/H_{t-4})$. A simulated value reconstructs the index by $H_t=H_{t-4}\exp(\pi_t^{\log}/100)$. Annual inflation is then $100((\sum_{q=1}^{4}H_{y,q})/(\sum_{q=1}^{4}H_{y-1,q})-1)$. For the bank model, half-year inflation compares the average index with the same half-year of the preceding year. Averaging the internal log changes cannot replace these definitions based on index levels.

\paragraph{Other variables.}
Unemployment and the long-term interest rate are period averages of the relevant quarterly observations. Equity and changes in residential property prices use ratios of the registered index levels. A comparison with an official scenario distinguishes changes in annual average indexes from changes between year-end indexes. The source definition determines which field is compared, and the program rejects a comparison lacking that correspondence. It never relabels one aggregation as another.

\paragraph{Annual official paths and half-year inputs.}
The bank validation uses official annual targets and an explicit assumption used for temporal disaggregation. For a positive variable whose target is the annual average $A_y$, let $P_{y-1,4}$ be the quarterly level at the end of the preceding year. Choose the unique positive number $g_y$ satisfying $P_{y-1,4}(g_y+g_y^2+g_y^3+g_y^4)=4A_y$, and set $P_{y,q}=P_{y-1,4}g_y^q$. The left side is strictly increasing for $g_y>0$, so the positive target determines a unique path within the year. This construction reproduces the official annual average without interpreting annual GDP growth as an endpoint return. An official scenario often states the annual growth rate $g^{\mathrm{a}}_{y}$ rather than the average level itself. The level targets then follow recursively from the last observed annual average as $A_{y}=A_{y-1}(1+g^{\mathrm{a}}_{y}/100)$, and the same recursion converts an official annual inflation path into targets for annual average indexes before the within-year construction runs.

An end-of-year index target instead uses the positive factor $g_y=(P_{y,4}/P_{y-1,4})^{1/4}$. Annual unemployment and interest-rate targets are held constant within the year. These assumptions are recorded as model inputs, not as observed official quarterly paths. Half-year variables are then calculated from the resulting quarterly levels using the same functions used for estimation. The input registry must supply the previous year's levels and period definitions before this operation can run.

\subsection{The Macro-Financial Model}

The model is a quarterly vector autoregression for a common euro area geography. Its six variables are the quarterly log changes in real GDP, equity prices, and residential-property prices, together with year-on-year log HICP change, unemployment, and the long-term interest rate. If $x_t$ collects these transformed series, the specification is
\begin{align}
 x_t=c+A_1x_{t-1}+A_2x_{t-2}+u_t,
 \label{eq:varscenario}
\end{align}
where $c$ is an intercept vector, $A_1$ and $A_2$ are coefficient matrices, and $u_t$ is the reduced-form innovation.

The sample begins in 2000 and ends at the latest common quarter supported by releases available on December 31, 2024. At least 80 complete quarterly observations are required. The endpoint and actual observation count are outputs of the selection based on dated data and are not fixed in advance to the fourth quarter of 2024. The program checks stability before drawing 20,000 trajectories by resampling entire residual vectors, thereby preserving contemporaneous dependence across the six variables.

Simulation covers the interval from the estimation endpoint through the fourth quarter of 2027. Quarters between the estimation endpoint and the scenario horizon are simulated along with the subsequent path. Annual 2024 denominators therefore combine eligible observed levels and simulated missing levels within each trajectory when needed. Every value retains its observed or simulated status. Annual paths for 2025--2027 are calculated only after those levels have been assembled.

Range checks use ordinary percentage changes or percentage levels under the definitions above. The configured annual intervals are $[-20,20]$ for GDP growth, $[-5,20]$ for inflation, $[0,30]$ for unemployment, and $[-2,20]$ for the long-term rate. Equity and changes in residential property prices use $[-80,50]$ and $[-60,40]$, respectively. Each variable's exact aggregation accompanies its bound.

The unconditional severity features use the cumulative change in GDP levels and the relevant price indexes over the scenario horizon, together with peak inflation, unemployment, and interest rates. These features are standardized over the common simulation set. Declines in the features based on level changes and increases in the three peak features carry equal adverse weights. The most adverse decile is clustered into three groups, and the most severe member of each group is selected. Conditions for individual risks use Equation~\ref{eq:restrictionseverity} instead.

\subsection{Bank Models and External Validation}

The bank application specifies a dynamic panel quantile regression with fixed effects, FE-QAR, and a dynamic linear model with fixed effects, FE-OLS, following the component approach of \citet{Covas2014stresstestingus}. Both use the eligible 2016--2024 EBA Transparency Exercise releases and the same accounting components. Credit impairment and pre-provision net revenue are modeled alongside residual charges, capital adjustments not explained by profit, and log growth in risk-weighted assets (RWA).

Let $i$ index banks, $t$ half-years, and $j$ accounting components. Let $y_{i,t}^{(j)}$ be a component rate, and let $b_{i,t-1}$ contain the lagged CET1 ratio and log RWA. The macroeconomic vector $x_{i,t}$ contains the current half-year scenario values for the bank's mapped country. The conditional quantile specification is
\begin{align}
\begin{aligned}
 Q_{\tau}\left(y_{i,t}^{(j)}\mid y_{i,t-1}^{(j)},b_{i,t-1},x_{i,t}\right)
 ={}&\alpha_i^{(j)}+\mu_{\tau}^{(j)}+\rho_{\tau}^{(j)}y_{i,t-1}^{(j)}\\
 &+\beta_{\tau}^{(j)\prime}b_{i,t-1}+\gamma_{\tau}^{(j)\prime}x_{i,t}.
\end{aligned}
 \label{eq:bankfeqar}
\end{align}
Here, $\alpha_i^{(j)}$ is the bank fixed effect, $\mu_{\tau}^{(j)}$ the intercept for quantile $\tau$, and the remaining coefficients multiply the lagged component, bank controls, and current scenario inputs. Conditioning explicitly includes $x_{i,t}$ because the projection treats the macroeconomic scenario as supplied information. FE-OLS uses the same regressors in a conditional mean equation.

Historical features and evaluation features use the same country mapping and temporal transformations. A euro area history must not be substituted for a estimation series for that country while variation by country is introduced only at evaluation. A bank requires at least eight complete observations, and a lag is formed only for consecutive observed half-years. December cumulative flows are converted to second-half flows, and all component rates use RWA at the beginning of the half-year.

FE-QAR uses 19 quantiles from 0.05 through 0.95 and a penalty on the fixed effects of one. Its joint objective is solved as a sparse linear program using the HiGHS interior-point method. FE-OLS uses the within transformation. Both specifications generate 25,000 projection paths per bank and supplied country scenario. Their parameter and input snapshots are recorded before the external comparison.

Let $n_{i,t}$, $\ell_{i,t}$, $o_{i,t}$, and $a_{i,t}$ denote percentage rates for pre-provision net revenue, credit losses, residual charges, and capital adjustments. If $g_{i,t}$ is 100 times log RWA growth, the common accounting recursions are
\begin{align}
 C_{i,t}&=C_{i,t-1}+\frac{n_{i,t}-\ell_{i,t}-o_{i,t}+a_{i,t}}{100}\operatorname{RWA}_{i,t-1},
 \label{eq:cet1capital}\\
 \operatorname{RWA}_{i,t}&=\operatorname{RWA}_{i,t-1}\exp\left(\frac{g_{i,t}}{100}\right),
 \label{eq:rwarecursion}\\
 \operatorname{CET1Ratio}_{i,t}&=100\frac{C_{i,t}}{\operatorname{RWA}_{i,t}}.
 \label{eq:cet1ratio}
\end{align}
Here, $C_{i,t}$ is CET1 capital. Model outputs must satisfy positive RWA and the stated accounting identities before the ratio is used.

External validation conditions all models on the same official December 2024 capital and RWA starting values and the same official adverse country scenario. Those evaluation inputs are kept separate from the historically available inputs used for generation. Any bridge from an earlier observed bank state to December 2024 retains its modeled status. A bank analysis under a generated scenario uses only bank information eligible at the historical cutoff and cannot import later starting values from the validation data.

The validation thresholds are mean absolute error in CET1 depletion no greater than two percentage points, Spearman correlation at least 0.5, and interval coverage at least 0.75. Comparisons with a no-change calculation use 2,000 bootstrap replications by country cluster. All three conditions must pass before a model supplies bank conclusions under generated scenarios. Agreement with the official hypothetical stress calculation is the target of this exercise, rather than prediction of realized future losses.

\section{Evaluation Measures}

The evaluation distinguishes recovery of economic relations from successful numerical execution. This section defines the relevant measures and their comparison units without treating the execution record as an measure of economic performance.

\subsection{Relations and Paths}

Coverage of candidate pairs records whether both variables of a benchmark relation occur in the eligible text considered for extraction. It is an upper bound on what a particular candidate screen could extract, not evidence that the text supports a direction or sign. Relation coverage checks whether the extracted graph contains the same ordered endpoints and sign as a benchmark relation. Benchmark match fraction divides the number of matching distinct relations by the number of extracted distinct signed relations. It does not classify unlisted relations as economically false.

Every benchmark path must satisfy continuity of consecutive relations before scoring. Endpoint path availability asks whether at least one admissible directed path connects the specified endpoints. Complete sequence presence asks whether all relations of one connected benchmark variant are present. Exact selected-path recovery additionally requires the predicted sequence to equal that variant, with no extra or misplaced relation. Ordered precision and recall use longest-common-subsequence overlap, with zero assigned when the predicted path is absent.

The measures based on all paths and the five highest-ranked paths report the best score attainable in their respective candidate sets. Their difference identifies loss caused by ranking before LLM selection. The primary unit is the initiating risk. Measures are shown for each of the three risks and averaged descriptively, without treating channels within a risk as independent replications.

\subsection{Numerical Scenarios and Bank Outcomes}

Scenario measures include directional violations, annual range checks, the number of distinct selected simulations, and sensitivity to the three priority weighting rules. Every comparison requires common transformation identifiers and units. Annual GDP growth and an endpoint GDP change are separate quantities, as are log returns and ordinary percentage changes.

Bank evaluation reports mean absolute error and root mean squared error in CET1 depletion, together with rank correlation, top-quartile recall, and interval coverage. Resampling by country cluster preserves the grouping of banks exposed to a common official macro path. The no-change comparator predicts zero depletion from the same starting state. All quantities compare calculations under the official adverse scenario rather than forecasts with realized future bank outcomes.

\subsection{Reports and Execution Records}

Completeness of source locations records attachment of a retrievable document identifier and exact location. Linkage to a model run records whether a numerical claim identifies a completed calculation. Numerical recomputation reads the referenced value and applies the stated transformation and rounding rule. Equality uses the recorded numerical tolerance, with $10^{-10}$ for the unrounded comparison between the unrounded claim and its source. Required metadata must agree as well as the number.

The test design separates positive controls from invalid inputs. Positive controls include legitimate signed values and valid numerical tokens. Invalid inputs include unknown claim identifiers, altered output hashes, wrong units, misplaced dates, and quantities written as words. Each valid case must remain accepted, and each invalid case must be rejected by both the checks during generation and after execution. Independent arithmetic examples test log conversion, annual GDP aggregation, HICP reconstruction, and disaggregation from annual to quarterly values.

Report completeness and numerical record consistency are distinct from economic support for the prose. Completion counts use all scheduled repetitions, and agreement conditional on completion is reported alongside agreement over all repetition pairs. A stopped repetition can have an internally consistent audit record without having produced a complete analysis.

\section{Date Fields Used in the Application}

The historical information cutoff concerns when a value or statement was available, not only the period it describes. A document published in late 2024 can contain earlier observations, and a release in 2025 can first publish or subsequently change a 2024 value. The design therefore records the publication date, observation period, and data vintage separately.

\subsection{Source Eligibility}

Generation evidence must have been published by December 31, 2024 and belong to an allowed empirical role. Historical official scenarios are reserved for development, while the official 2025 scenarios and bank outcomes are evaluation material \citep{ESRB2023macrofinancial,ESRB2025macrofinancial,BOE2025variablepaths,EBA20252025eu}. Role labels apply at the section or chunk level when a report combines ordinary risk discussion with a stress scenario reserved for development. For example, the chapter reporting the Bank of England's 2024 desk based stress test is excluded from generation even though it appears in a financial-stability report published before the cutoff \citep{BOE2024financialstabilitynovember}.

Let $G=(\mathcal{N},\mathcal{A})$ be the stored graph, and let $d_r^{\mathrm{pub}}$ denote the publication date of relation record $r$. Write $\mathcal{A}_{\mathrm{gen}}$ for the records assigned an allowed generation role. For an information date $t$, the eligible edge set is
\begin{align}
 \mathcal{A}(t)=\{r\in\mathcal{A}_{\mathrm{gen}}:d_r^{\mathrm{pub}}\leq t\text{ and the requested scope is supported}\}.
 \label{eq:asofgraph}
\end{align}
The corresponding nodes and edges form the retrieval graph. An unknown publication date causes exclusion. A recorded observation sample describes the basis of an empirical finding; it does not have to extend into the future scenario horizon. An explicit validity interval is applied only when the source states that a rule or relation is valid during that interval.

\begin{figure}[!htbp]
\centering
\begin{tikzpicture}[
 box/.style={draw,rounded corners,align=center,minimum height=1.0cm,text width=3.0cm},
 arr/.style={-{Latex[length=2mm]},thick}]
\node[box] (gen) {Eligible evidence\\Released by the cutoff};
\node[box,right=0.65cm of gen] (cut) {Information cutoff\\December 31, 2024};
\node[box,right=0.65cm of cut] (eval) {Evaluation only\\Official 2025 records};
\draw[arr] (gen) -- (cut);
\draw[arr] (cut) -- (eval);
\node[box,below=0.8cm of cut] (dev) {Development only\\Historical official scenarios};
\end{tikzpicture}
\caption{Source roles in the design with a historical information cutoff. Development and evaluation material are excluded from generation.}
\label{fig:date-split}
\end{figure}

\subsection{Observed Values and Data Vintages}

Each observation records its reference period and the release date of the particular vintage used. For a provider series and reference period, the input selector chooses the latest release no later than the cutoff. Transformation and resampling occur after this choice. The registry also identifies the geographic membership and seasonal adjustment convention so that nominally similar regional series are not combined under one variable definition.

The preliminary flash estimate of fourth-quarter 2024 GDP was released on January 30, 2025 \citep{Eurostat2025gdpflash}. It is therefore excluded from a December 31, 2024 information set even though its quarter can be indexed by October 1, 2024. A file downloaded later establishes the value in that downloaded version, not the value available at the historical cutoff.

The numerical estimation sample ends at the latest common quarter supported by eligible releases for all required series. Later quarters needed to reach the scenario horizon are simulated from that endpoint. Missing archival releases stop the experiment based on dated data; a current download cannot silently supply them. Tests include observations released just before and just after the cutoff, multiple releases for the same period, and inconsistent regional membership.

\begin{table}[!htbp]
\centering
\caption{Dates retained by the analysis records.}
\label{tab:temporal-fields}
\begin{tabularx}{\linewidth}{p{0.27\linewidth}Y}
\toprule
Record & Temporal fields \\
\midrule
Source & Publication date and retrieval time \\
Relation & Source date, sample period, stated validity interval \\
Observation & Reference period, release date, vintage \\
Scenario & Information cutoff and projection horizon \\
Model specification & Estimation sample and parameter date \\
Model run & Start and end times, input versions \\
\bottomrule
\end{tabularx}
\end{table}

\section{Source Collection Details}

Table~\ref{tab:corpus-inventory} records the role labels assigned to files in the collected corpus. Eligibility for the planned experiment is determined at the source location after the content and date checks, so a label assigned to a file does not make every chunk in that file eligible.

\begin{table}[!htbp]
\centering
\caption{Corpus inventory.}
\label{tab:corpus-inventory}
\begin{tabularx}{0.86\linewidth}{Y r r r}
\toprule
Role & Files & Text chunks & File-level label \\
\midrule
Generation evidence & 28 & 3,149 & Generation \\
Model and retrieval methods & 17 & 865 & Excluded \\
Development references & 3 & 120 & Excluded \\
Evaluation references after the information date & 3 & 54 & Excluded \\
\midrule
Total & 51 & 4,188 & 28 generation \\
\bottomrule
\end{tabularx}
\end{table}

The files labeled for generation contain assessments of financial stability and macro-financial conditions. The main European documents in the collection are the \emph{Financial Stability Review} of the European Central Bank and the \emph{Financial Stability Report} of the Bank of England. Selected international sources include the International Monetary Fund \emph{Global Financial Stability Report}, publications of the Bank for International Settlements, and Basel Committee guidance \citep{ECB2024financialstabilitymay,ECB2024financialstabilitynovember,BOE2024financialstabilityjune,BOE2024financialstabilitynovember,BIS2024annualeconomic,BCBS2013principlesfor,BCBS2018stresstesting}. These publications identify risks that the issuing institutions considered material at a particular time.

Research papers and central-bank working papers have a different role because they provide evidence about the transmission mechanisms and modeling approaches needed for scenario construction. Citation counts alone do not determine whether a study supports the relation required by a current risk query. Instead, findings are used only when the study design and empirical scope agree with the analysis request.

The structured files include Bank of England stress scenario workbooks and Systemic Risk Survey appendices. The workbooks preserve variable metadata and historical observations alongside the baseline and adverse projections, whereas the survey appendices provide repeated measurements of risks cited by market participants. Before a table is converted into text, the parser records each cell's workbook location and labels together with its formula or value, so that labels and scenario conditions can be checked against the extracted value.

\subsection{Composition of the Source Collection}

Table~\ref{tab:generation-sources} reports the institutional composition of the 28 files labeled for generation, in which Bank of England and European Central Bank publications account for most European risk assessments. International institutions provide information about global macro-financial and regulatory relations, while Federal Reserve sources are used mainly for public stress model documentation.

\begin{table}[!htbp]
\centering
\caption{Institutions represented in the generation collection.}
\label{tab:generation-sources}
\begin{tabular}{lr}
\toprule
Institution & Files \\
\midrule
Bank of England & 14 \\
European Central Bank & 6 \\
Federal Reserve Bank of New York & 3 \\
International Monetary Fund & 2 \\
Bank for International Settlements & 2 \\
Board of Governors of the Federal Reserve System & 1 \\
\midrule
Total & 28 \\
\bottomrule
\end{tabular}
\end{table}

The larger Bank of England subcorpus can influence term frequency in document retrieval, so Table~\ref{tab:generation-sources} is needed when interpreting the retrieval results. Official reports and research papers also serve different purposes: a policy assessment can establish which risks were prominent at a date, whereas an empirical study can provide evidence for a specified economic relation. The graph retains the source type and publication date for every accepted relation.

\section{Risk Information Available by December 2024}

\begin{figure}[!htbp]
\centering
\includegraphics[width=0.92\linewidth]{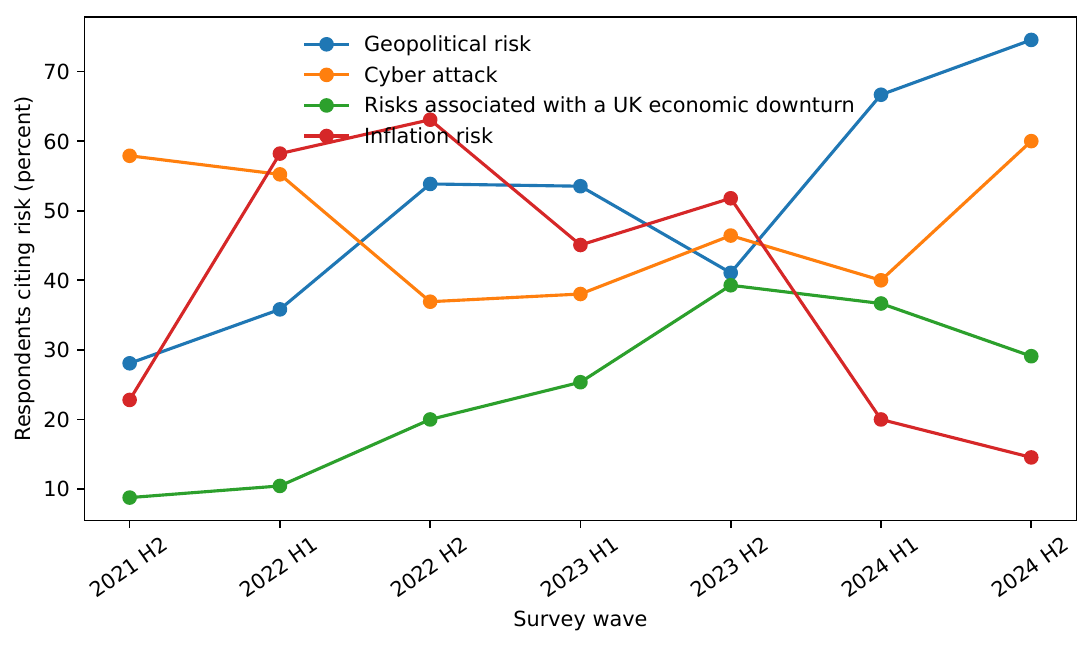}
\caption{Selected risks cited in the Bank of England Systemic Risk Survey. Values are percentages of respondents citing each risk.}
\label{fig:boe-risk-probabilities}
\end{figure}

The repeated Bank of England Systemic Risk Survey provides a time-indexed measure of risks cited by market participants. Figure~\ref{fig:boe-risk-probabilities} reports four risks from the second half of 2021 through the second half of 2024. The figure is constructed from the published data appendices and characterizes the information available before the principal information date \citep{BOE2024systemicrisk}.

Geopolitical risk rises from 28.1 percent in the second half of 2021 to 74.5 percent in the second half of 2024, when cyberattack risk reaches 60.0 percent. The share citing a United Kingdom economic downturn increases through 2023 and is 29.1 percent in the survey in the second half of 2024. By contrast, inflation risk peaks above 60 percent in 2022 and declines to 14.5 percent by the second half of 2024. These observed percentages describe the information available before the evaluation date rather than conclusions produced by a quantitative model. Their change over time also shows why a date restriction is needed: a source collection without one could overemphasize inflation assessments from 2022 and understate the later rise in geopolitical and cyber risks.

The survey identifies perceived sources of risk, but it does not contain the complete channels that connect an initiating shock to economic conditions and bank capital. Consequently, the survey contributes observed values to risk discovery, while research papers and financial stability reports provide the additional relations required for scenario construction.

\section{Implementation Records and Diagnostic Tests}

The saved implementation records support inspection of source use and numerical references. The diagnostics specified here distinguish those checks from the economic evaluation and define the tests required for the dated data and report specifications.

\subsection{Graph and Source Records}

A displayed relation identifies its source document, location, quotation, endpoints, and sign. The analysis records the graph condition alongside selected paths and later model requests. A source-contribution calculation reconstructs the same graph after removing one document, and its output names the measure being recomputed. A missing path after removal identifies dependence on that source for that measure. It does not measure the semantic quality of every remaining relation.

The saved scenario records generated with G0 and G2 construction records are inspected separately. An complete comparison using G2 must create new path, restriction, numerical, and report records that all identify G2. Changing a label on a result generated from G0 would not satisfy this requirement.

\subsection{Numerical Input Tests}

The dated-input tests operate before estimation. A reference period that precedes the cutoff is rejected when its selected release occurs later. Multiple vintages must resolve to the latest eligible release, and source geography must remain constant under the registered definition. Removing a required vintage causes a failure rather than selection of a different data source.

Transformation tests compare positive levels with their exact log changes and reconstruct the same levels from those changes. Annual GDP tests compare ratios of four-quarter totals, and half-year tests compare the prescribed two-quarter totals. HICP tests reconstruct quarterly indexes from year-on-year log changes before calculating inflation. Disaggregation tests recover the supplied annual averages or endpoint targets from the quarterly path. Test values cover zero changes, large contractions, increases, and invalid nonpositive levels. Such tests validate formulas without claiming that a fitted economic model has passed external validation.

\subsection{Report and Model-Call Tests}

A report test must establish both that allowed numerical content is preserved and that prohibited content is rejected. Dedicated cases include percentages written as words, money amounts within the numerical range of calendar years, changes in basis points, repeated tokens, and unknown identifiers. The temporal field admits only the permitted time metadata. Numerical fields require a reference to a model output and cannot be populated directly by a LLM response.

Model-call tests verify that a numerical conclusion cannot be released when the required call was omitted, failed, or returned incompatible metadata. Altering a model-run identifier, output file, row, column, or transformation must invalidate the affected claim. These tests are applied to every retrieval representation with the same registered model inputs. Their outcomes are reported separately from the completion statistics of the recorded agent configurations.

\section{Fields Used in the Analysis}

This section specifies the records required by the analysis algorithm. It distinguishes the implemented report interface based on registered numerical tokens from the additional fields and tests specified for numerical evaluation. Acceptance of a structured record establishes that its fields satisfy their schema; whether a source supports an economic interpretation requires a separate check.

\subsection{Relations Reported in Sources}

A source relation identifies its two variables and direction together with its attached evidence. The evidence class records the basis for interpretation and is separate from the relation verb returned by the extractor. An unknown class or scope attribute is retained explicitly until the proposed use can be justified.

\begin{table}[!htbp]
\centering
\caption{Required information for a source relation.}
\label{tab:source-relation-fields}
\begin{tabularx}{\linewidth}{p{0.25\linewidth}Y}
\toprule
Field group & Content \\
\midrule
Identity & Relation and source and location identifiers \\
Relation & Endpoints, direction, and sign \\
Evidence & Quotation, evidence class, and method \\
Applicability & Jurisdiction, sector, regime, and horizon \\
Review & Decision and unresolved attributes \\
\bottomrule
\end{tabularx}
\end{table}

An extraction record must retain the original passage and its location. A recorded verb such as ``causes'' does not establish causal identification. Relations used to constrain a numerical scenario require a supported interpretation under the request's scope. Contextual relations can remain available to the report without being promoted to numerical restrictions.

\subsection{Observed Values}

The observation schema stores the value and its definition before that value is supplied to a model. Its definition identifies the unit, transformation, frequency, geographic coverage, and source series. Reference periods are stored separately from release dates and the archived data vintage. A index for the start of the quarter is not used as a substitute for the release date. Positive GDP and price index levels are retained wherever the output transformations require them.

The numerical registry rejects a requested observation when its release date or vintage is unavailable, or when its coverage conflicts with the model specification. Model output uses a separate schema because it depends on a completed scenario request and a particular estimation or calibration snapshot.

\subsection{Scenario Specifications}

The scenario specification connects an initiating disturbance to accepted channels and requested outputs. Each restriction identifies its variable, direction, reference value, and applicable horizon. It also identifies the underlying relation record or the analytical rule under which the restriction is imposed. Supporting and opposing source locations remain separate.

The evaluation record includes the graph condition and source eligibility manifest used to form the request. It records connected paths as ordered edges; a collection of parallel mechanisms is represented as several paths. A structural test verifies adjacent endpoints before the paths can be used or evaluated.

\subsection{Inputs to Quantitative Models}

The model input specification identifies the required variables and their definitions together with a data snapshot and an executable model entry point. Compatibility requires agreement in the output horizon, temporal aggregation, geographic coverage, and transformation. The runner checks those definitions before executing the requested function.

For the bank application, estimation and scenario inputs use the same country mapping and accounting conventions. An observed bank state and a model calculation that bridges to the scenario starting date have different status. The record cannot replace the former with later official starting values while continuing to label the analysis as historically available.

\subsection{Model Run Logs}

A model run record links the input request to its execution and output. Table~\ref{tab:model-run-contract} gives the required groups of information.

\begin{table}[!htbp]
\centering
\caption{Information required to reproduce a model run.}
\label{tab:model-run-contract}
\begin{tabularx}{\linewidth}{p{0.25\linewidth}Y}
\toprule
Field group & Content \\
\midrule
Identity & Run, scenario, and model identifiers \\
Inputs & Request and input hashes \\
Specification & Code, parameters, and environment \\
Execution & Start, end, status, and diagnostics \\
Outputs & File hash and locations of output points \\
Dependencies & Preceding runs and evaluation role \\
\bottomrule
\end{tabularx}
\end{table}

Each output point retains its variable, unit, transformation, and horizon. When the model provides uncertainty information, the record identifies the interval definition or the simulated distribution. Bank outcomes additionally retain the accounting components from which the reported capital ratio is calculated.

\subsection{Tests for Numerical Conclusions}

An \OutputTest{} specifies a numerical claim's required variable, unit, horizon, and model before execution. It also fixes the transformation, diagnostics, rounding rule, and tolerance used to compare the report with its source calculation. After execution, the same test links the model run and output point to its pass or fail decision. Several numbers in one sentence require separate numerical records.

The test design distinguishes agreement with a stored output from correctness of the transformation that created that output. Unit tests use independently calculated examples for log changes, growth in the price index, and period aggregation. Integration tests check that a value's metadata remain unchanged between the model input, result table, and report. These tests specify numerical properties without prescribing an unknown scenario outcome.

\subsection{Fields Used in Graph Queries}

The graph query identifies the initiating variable, target outcome, maximum path length, and source eligibility conditions. The program enumerates candidates under these fields before the language model selects among them. The selected path must be one of the enumerated candidates and must retain the graph condition that generated it.

The maximum path length and manifest of eligible sources are part of the request, not editable LLM output. A path with an unknown endpoint, unsupported sign, or inconsistent scope is rejected before numerical execution. An empty admissible set stops the corresponding risk program.

\subsection{Report Sentences}

The report specification separates an attributed qualitative statement from a numerical conclusion produced by a model and from temporal metadata. A qualitative statement links to its attached source locations. A numerical conclusion identifies a passed test and a completed model run. Dates use dedicated fields rather than exemptions for any integer that resembles a year.

The recorded implementation uses registered numerical tokens. Resolving those tokens checks the registered values but does not exclude quantities expressed elsewhere in free text. The numerical evaluation design tests quantities written as words and integers accompanied by units, including money and basis-point changes. Checks during generation and after execution must reject the same invalid forms. Raw observations may appear in descriptions of inputs, with their source and definition, but cannot be substituted for a conclusion produced by a model.

\section{Comparison Design}

This section records the contrasts and execution records required for the numerical evaluation. It separates that design from the operational outcomes of the saved multi-agent runs.

\subsection{Primary Contrasts}

Graph construction is compared under matched source chunks and extraction resources. Verification is evaluated by applying it to the same candidate set, so its effect is not conflated with a change in retrieval coverage. The complete comparison then fixes the numerical model and reports the graph condition used by every downstream operation.

\begin{table}[!htbp]
\centering
\caption{Primary comparisons and their evaluation units.}
\label{tab:primary-contrasts}
\begin{tabularx}{\linewidth}{p{0.38\linewidth}Yp{0.19\linewidth}}
\toprule
Comparison & Primary outcome & Unit \\
\midrule
Decomposition into short statements & Supported relation coverage & Source chunk \\
Candidate verification & Support and rejection rates & Proposed relation \\
Graph path selection & Connected relation match & Initiating risk \\
Retrieval method & Source and model linkage & Report \\
Scenario construction & Directional violations & Initiating risk \\
Bank model specification & Error in CET1 depletion & Country cluster \\
\bottomrule
\end{tabularx}
\end{table}

The graph benchmark is structurally validated before scoring. A single path cannot contain disconnected edges, and occurrences of a shared edge are distinguished from distinct signed relations. The three initiating risks are reported separately rather than treating their constituent channels as independent replications.

\subsection{Configuration}

The information cutoff is December 31, 2024, and the scenario horizon is 2025--2027. Eligibility is determined by release date and vintage, while the estimation endpoint follows from the available observations. The registered graph query permits at most five relations per path and five candidate paths. Each retrieval condition receives the same total limit on evidence size in bytes and the same output schema.

The recorded model identifier is \GPTFourOSnapshot{}. The planned comparison fixes the model identifier, temperature, schema for structured responses, and request budget before execution. A different or unavailable model ends that execution rather than silently changing the comparison. Automatic retries are disabled, and any explicitly permitted repair is counted as an additional request.

The numerical design uses a VAR with two lags and 20,000 residual bootstrap simulations. Both bank specifications use 25,000 projection paths and are compared with a no-change calculation using 2,000 bootstrap replications by country cluster. The validation thresholds and accounting equations are specified in the section on quantitative models. Those thresholds remain fixed during evaluation.

\subsection{Evaluation Records}

Before the evaluation files are opened by a run, the execution program stores hashes of the manifest of eligible sources, model inputs, configuration, generated graph, scenario restrictions, and numerical outputs. The evaluation record stores the benchmark and official comparison data separately. Because the benchmark has already been inspected during method development, this computational separation is not presented as an untouched prospective test.

Source attachment, semantic support, recovery of connected paths, and numerical consistency remain separate outcomes. A report can retain a valid source identifier while overstating what the cited passage supports. Similarly, a recorded number can equal a stored model cell while inheriting an incorrect transformation. Independent tests address those different questions rather than combining them into a single acceptance percentage.

\subsection{Reasons for Stopping}

The run stops when a required input is missing or has the wrong hash, a response violates its schema, no admissible path exists, or a numerical output violates a required identity. A risk program also stops before any model call when the restrictions derived from its accepted channels are fewer than the model request schema requires; the derivation record retains every assessed path, its composed direction, and the reason for the stop. Unsupported data vintages and incompatible geographic coverage also stop the affected calculation. A failure remains visible in the execution record and is not replaced by an old result or a newly selected model.

\subsection{Information Stored for Reproduction}

Every source file has a recorded SHA-256 hash. API records retain the request, raw response, parsed object, and requested and returned model identifiers together with token use and content hashes. Quantitative records retain the random seeds and the data, code, configuration, and output hashes. Analysis notebooks generate empirical tables and figures from those numerical records. The evaluation design requires its own execution records before numerical comparisons can be reported.

\section{Generated Scenario Reports}
\label{app:generated-reports}

This appendix reproduces the reports generated in the recorded Run 7 execution as the output of the algorithm in Section~\ref{sec:agent}. Each report was produced by the report agent from the accepted evidence and its completed model run, passed the admissibility checks, and is rendered here by resolving every registered numerical token from the claims of that model run; a directional sentence carries the source attached to its cited relation. Each report is set in a frame, and the framed text is the report. The registered values are identical across the complete repetitions, whose reports differ only in wording, and the reports of the first repetition that completes both derivable risk programs are reproduced verbatim. Market illiquidity has no report because its program stops before any model call.

\subsection{Property-Market Correction}

\begin{framed}
\noindent\emph{Impact of Higher Financing Costs on European Property Markets}

Higher financing costs are contributing to a broad correction in European property markets. (ECB FSR Nov 2023, p.~31) In 2025, house prices are expected to return 5.170\%. This decline in house prices is accompanied by a projected real GDP growth of 0.775\% in the same year. As financing conditions tighten, house prices are expected to continue their downward trend with a return of -2.527\% in 2026. (ECB FSR Nov 2023, p.~31) The unemployment rate is projected to be 7.843\% in 2026, reflecting the broader economic challenges. By 2027, house prices are expected to return -5.379\%, continuing the correction trend. The long-term interest rate is projected to be 5.442\% in 2027, maintaining pressure on financing conditions. Overall, these factors contribute to a challenging environment for the European property market, driven by higher financing costs and weaker income.
\end{framed}

\subsection{Trade Fragmentation}

\begin{framed}
\noindent\emph{Impact of Trade Fragmentation on Global Economic Indicators}

The escalation of geopolitical tensions and trade restrictions is expected to fragment global trade, leading to increased economic uncertainty. (ECB FSR Nov 2023, p.~36) This heightened uncertainty is likely to increase market volatility, affecting various economic indicators. (BoE FSR Dec 2022, p.~14) In 2025, real GDP growth is projected to be -1.142\%, reflecting the adverse effects of tighter financing conditions. (ECB FSR Nov 2023, p.~6) Accompanying this, the unemployment rate is expected to reach 6.152\%, indicating a challenging labor market. Equity returns in 2025 are anticipated to be -18.131\%, as market volatility impacts equity prices. (IMF GFSR Apr 2024, p.~18) House price returns for the same year are projected at 5.091\%, reflecting the broader economic slowdown. Inflation in 2025 is expected to be 2.048\%, driven by increased geopolitical risks. (ECB FSR May 2024, p.~96) By 2026, real GDP growth is projected to slow further to 0.595\%, continuing the trend of economic contraction. Inflation is expected to rise to 3.953\%, as financing conditions remain tight. (IMF GFSR Apr 2024, p.~11) The unemployment rate in 2026 is projected to be 7.144\%, reflecting ongoing labor market challenges. Equity returns for 2026 are anticipated to be -46.483\%, as market conditions remain volatile. (IMF GFSR Apr 2024, p.~18) In 2027, real GDP growth is expected to be -17.491\%, indicating a continued economic slowdown. The unemployment rate is projected to rise to 8.107\%, further highlighting labor market difficulties. House price returns in 2027 are expected to be -0.927\%, as economic pressures persist. Equity returns for 2027 are projected at -56.517\%, continuing the trend of market volatility. (IMF GFSR Apr 2024, p.~18) Inflation in 2027 is anticipated to be 3.173\%, as geopolitical tensions continue to exert upward pressure. (ECB FSR May 2024, p.~96)
\end{framed}

\bibliography{arXiv2.bbl}

\bibliographystyle{tmlr}

\end{document}